\documentclass[useAMS,usenatbib]{mn2e}
\usepackage{graphicx}
\usepackage{epsf}
\usepackage{amsmath}
\newcommand{\hangpar}{\noindent\hangindent.1in}
\newcommand{\teff}[1]{$T_{\rm eff}$}
\newcommand{\vsini}[1]{$v\cdot\sin(i)$}
\def\cm1{$\rm cm^{-1}$}
\def\kms{$\rm km\,s^{-1}$}
\def\DE{D\kern-0.75em \raisebox{1.0pt}{=}\ }

\def\Sum{N_{\rm tot}}
\def\Rh{\rule{15.0pt}{0.0pt}}
\def\Rv{\rule[-0.1in]{0.0pt}{20.0pt}}

\title[HD 65949: Rosetta Stone or Red Herring?]
{HD 65949: Rosetta Stone or Red Herring
\thanks{Based on observations
obtained at the European Southern
Observatory, Paranal and La Silla, Chile
(ESO programmes
076.D-0172(A), 081.D-0498(A)), HARPS data
obtained during engineering nights, and
at the Complejo Astron\'omico El Leoncito.}}
\author[C. R. Cowley, S. Hubrig, P. Palmeri, et al.]
{C. R. Cowley${^1}$
\thanks{E-mail: cowley@umich.edu},
S. Hubrig${^2}$,
P. Palmeri${^3}$,
P. Quinet${^{3,4}}$,
\'{E}. Bi\'{e}mont$^{3,4}$,
\newauthor G. M. Wahlgren$^{5,6}$,
O. Sch\"{u}tz$^7$,
and J. F. Gonz\'{a}lez${^8}$   \\
$^{1}$Department of Astronomy, University of Michigan,
   Ann Arbor, MI 48109-1042, USA\\
$^{2}$AIP, An der Sternwarte 16, 14482 Potsdam, Germany \\
$^{3}$Astrophysique et Spectroscopie, Universit\'{e}
de Mons, B-7000, Belgium  \\
$^{4}$IPNAS, Universit\'{e} de Li\`{e}ge, Sart Tilman
B15, B-4000 Li\`{e}ge, Belgium \\
$^{5}$Catholic Univ. of America, 620 Michigan Ave NE,
Washington, DC 20064, USA \\
$^{6}$NASA Goddard Space Flight Center, Code 667,
Greenbelt, MD 20771, USA \\
$^{7}$European Southern Observatory, Casilla 19001,
Santiago 19, Chile\\
${^8}$Instituto de Ciencias Astron\'omicas, de la Terra y
del Espacio, Casilla 467, 5400
San Juan, Argentina
}
\begin{document}

\date{Accepted . Received ; in original form }

\pagerange{\pageref{firstpage}--\pageref{lastpage}} \pubyear{2008}

\maketitle

\label{firstpage}

\begin{abstract}
HD 65949 is a late B star with exceptionally strong Hg II
$\lambda$3984, but it is not a typical HgMn star.  The
Re II spectrum is of extraordinary strength.
Abundances, or upper limits are derived here
for 58 elements
based on a model with $T_{\rm eff} = 13100$K, and
$\log(g) = 4.0$.
Even-Z elements through nickel show minor deviations
from solar abundances.  Anomalies among the odd-Z elements
through copper are mostly small.  Beyond the iron
peak, a huge scatter is found.
Enormous enhancements are found for the elements rhenium
through mercury (Z = 75--80).  {\bf We note the presence of
Th III in the spectrum.}
The abundance pattern of the heaviest elements
resembles the N=126 r-process peak of solar material, though
not in detail.  An odd-Z anomaly appears at the triplet
(Zr\,Nb\,Mo), and there is a large abundance jump between
Xe (Z = 54) and Ba (Z = 56).  These are signatures of
chemical fractionation.

We find a
significant correlation of the abundance excesses with
second ionization potentials for elements with Z $>$ 30.
If this is not a red herring (false lead), it indicates
the relevance of photospheric or near-photospheric processes.
Large excesses
(4-6 dex) require diffusion from deeper layers with the
elements passing through a number of ionization stages.
That would make the correlation with second ionization
potential puzzling.
We explore a model with
mass accretion of exotic material followed by the
more commonly accepted differentiation by diffusion.
That model leads to a number of predictions which
challenge future work.

New observations confirm the orbital elements of
Gieseking and Karimie, apart from
the systemic velocity, which has increased.  Likely
primary and secondary masses are near 3.3 and 1.6 $M_\odot$,
with a separation of ca. 0.25 AU.

New atomic structure calculations are presented in two
appendices.  These include partition functions for the
first through third spectra of Ru, Re, and Os, as well
as oscillator strengths in the Re II spectrum.

\end{abstract}

\section{Toward an understanding of CP stars}

{\it \bf In situ} chemical separation, under gravitational and
radiative forces is accepted as the basic explanation of
abundance anomalies in upper main sequence, chemically
peculiar (CP) stars.
Nevertheless, there have been few breakthroughs
of the stature of arguments originally
posed by Michaud (1970).  Briefly, these were that the
anomalies appeared in the stable atmospheres of
slowly rotating stars with
radiative envelopes.  Additionally, the more abundant elements,
helium, carbon, nitrogen, and oxygen could have little
radiative support because their strong lines would be
saturated.  Time has not dimmed the relevance of that
insight.

Scientific breakthroughs often hinge on the location of
special cases, where the effects under consideration are large.
A code breaker is at a severe disadvantage when faced with a
brief message.  With a long message, it is more likely
that the regularities of a language will lead to
a decryption key.  We hope that the present study represents
a kind of longer message, and can serve as a Rosetta stone,
for an understanding of the more bizarre anomalies
seen in CP stars.  We provide information on more
elements (58)
than in a typical study of similar stars (20-30).  Additionally,
a number of the anomalies are very large.

The extensive analysis of Castelli and Hubrig (2004, CH04) provides
a guide for the present work.  Their study of the classical
HgMn star, HR 7143 (HD 175640),
reported abundances for 40 elements.  The
abundance anomalies are similar in some ways to those of
HD 65949, and dissimilar in
others.  It has been helpful to compare
results for the two
stars.  A detailed comparison with HR 7143
has been possible because of
the spectra posted on Castelli's (2009) web site.

The HgMn star $\chi$ Lup has also been the subject of intensive
study (cf. Wahlgren 2005, and many cited references therein).
The star is significantly cooler ($T_{\rm eff} = 10650$K)
than HD 65949 (ca. 13100K), and many important results
were obtained from {\it Hubble Space Telescope}
($HST$) observations for which there is no
comparable material for HD 65949.  We briefly discuss the
$\chi$ Lup abundances in the light of the present study.

\section{An unusual late-B spectrum}

Abt and Morgan (1969) noted the great strength of Hg {\sc ii}
$\lambda$3984 in the spectrum of HD 65949, and remarked
that it did not seem to be an HgMn star.  Hubrig et al.
(2006) reported a weak magnetic field which might indicate
a relation to the magnetic sequence of CP stars (Preston
1974).

More recent high-resolution ESO observations revealed a
truly unusual spectrum
(Cowley et al. 2006, Paper I, Cowley, Hubrig,
\& Wahlgren 2008, Paper II).  In addition to the possibly
record-setting strength of Hg {\sc ii} $\lambda$3984, along
with strong Pt {\sc ii}, lines of Os {\sc ii} and especially Re {\sc ii}
were numerous.  Osmium and rhenium have been investigated
in the ultraviolet spectrum of $\chi$ Lup (Wahlgren et al.
1997, Ivarsson et al. 2004), but the presence of lines
of these elements in ground-based spectra is
unusual.

{\bf The richness of the line spectrum is due not only to the
unusual abundances.  The lines are extremely sharp.
We estimate from spectral synthesis, that
$v\cdot\sin{i} = 0.5 \pm 0.5$ km\,s$^{-1}$.}

The present work is a
more complete abundance analysis of HD 65949, though
based primarily on equivalent widths.
For rich spectra, such as Fe {\sc  i} and {\sc ii}, we
obtained, hopefully, a sufficient number of measurements,
but did not attempt to analyze all possibly relevant features.
In this respect, the CH04 work is undoubtedly superior,
since the entire spectrum of HR 7143 was synthesized.
However, relatively small errors in the
present analysis are
less important than they might otherwise be, given the
large departures of many values from the standard (solar)
abundance distribution (SAD, e.g. Asplund, Grevesse,
Sauval, and Scott 2009).

HD 65949 is located in the young cluster NGC 2516, which is
known for more than a typical number of CP stars.  This
cluster also has an unusual number of X-ray sources
(Wolk et al. 2004).
These facts make it tempting to suggest that mass
transfer might be relevant for some aspect of the
anomalies.  This is an old idea, which Wahlgren et al.
(1995) remarked ``remains a distant alternative, but
possibly a collaborator to diffusion theory.''

\section{The atmosphere of HD 65949}

The effective temperature of HD 65949 is uncertain by
several hundred degrees.  The estimate used in Paper I,
$T_{\rm eff} = 13600$K, came from averaged Str\"{o}mgren and
H$\beta$ photometry (Hauck \& Mermilliod 1998),
and the calibration of
Moon and Dworetsky (1985) and implemented by Moon (1984).
We used a version of the Moon code kindly supplied by
Dr. B. Smalley.  For Paper {\sc ii}, we adopted a temperature 1000K
lower, which gave equal abundances for Fe {\sc i}, {\sc ii},
and {\sc iii}.  Here, the reasoning was that for abundances
it is more important to have the ionization correct
than the color temperature.  Since that work, we have
become more convinced of the plausible relevance of
stratification in the atmospheres of early stars.
When an element is non-uniformly distributed in a
photosphere, the apparent ionization temperature of
one element will not in general indicate the correct
degree of ionization of another.

A computer code kindly supplied by Dr. P. North (cf.
Kunzli et al. 1997) allows one to include an abundance
estimate in the calculation of $T_{\rm eff}$ and $\log(g)$.
Geneva photometry was obtained from
the online General Catalogue
of Photometric Data of Mermilliod, Hauck, \& Mermilliod (2007).
The code requires a reddening estimate, which we obtained
from measurements of the interstellar
Na {\sc  i }D$_2$ and K {\sc i} resonance lines, with the help of the
calibration of Munari and Zwitter (1997).  We adopted
$E(B-V) = 0.042$ (smaller than typical measurements
for NGC 2516, ca. 0.1; cf. references in van Leeuwen 2009).
Conversions
of reddening from the $UBV$ system they used to the Geneva
system were taken from Paunzen, Schnell, \& Maitzen (2006).
Dr. North's code calculates $T_{\rm eff}$ and $\log(g)$
for [Fe/H] values of
$-1$, 0, and +1.  We averaged the results for 0 and +1,
and obtained the adopted value $T_{\rm eff} = 13100$K.
This value is conveniently between that obtained from
Str\"{o}mgren photometry and that giving iron ionization
equilibrium.

\begin{figure}
\includegraphics[width=54mm,height=84mm,angle=270]{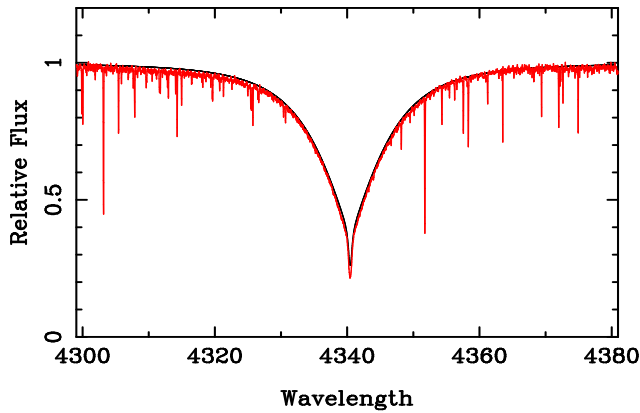}
 \caption{H$\gamma$ profile
for $T_{\rm eff} = 13100$K,
$\log(g)=4.0$.  The calculated profile is nearly obscured by
the observations (gray, red in online version).  The HD 65949
profile is from a HARPS (Mayor et al. 2003)
spectrum obtained on 31 March 2009.
No metal lines
were included in the calculation. \label{fig:hgamma}}
\end{figure}

Dr. North's code also gives surface gravity: $\log(g) = 4.2$.
However, our calculations were all made with $\log(g) = 4.0$.
Low Balmer profiles are relatively insensitive to the
effective temperatures considered here, but agree well with
the assumed $\log(g) = 4.0$ (Fig.~\ref{fig:hgamma}).

No stratification was assumed in any of the abundance
calculations.

\section{Spectra}
\label{sec:spectra}

ESO/FEROS spectra are discussed in Kaufer et al. (1999).
The useful
coverage was $\lambda\lambda$3603-9211, with a
resolving power
of 48000.  They were supplemented
by a UVES (Dekker et al. 2000) spectrum obtained on 4 August 2008
($\lambda\lambda$3258-4517, and 5655-9464).  The
resolving power in
the blue arm is 80000, and 110000 in the red.  Additionally,
we used
HARPS spectra obtained on 14 December 2008, 31 March 2009,
2 June 2009, and 3 June 2009.  The wavelength coverage is
$\lambda\lambda$3972-6911, with a resolving power of 120000.
Most quantitative measurements were made on the December 2008
spectrum, where the signal-to-noise (S/N) was
120 to 150.

{\bf Most abundances are based on
equivalent width measurements} carried out with Michigan
software, which fit a Voigt profile to the stellar features.
Equivalent widths for few hyperfine and helium profiles
were obtained from {\bf quadrilaterals} or triangles estimated to
have the same area as the more complex absorption profiles.

\section{Binarity}
\label{sec:binarity}

\begin{table}
\caption{New radial velocity measurements of HD 65949\label{tab:newrv}}
\begin{center}
\begin{tabular}{c c r r l}\hline
& {\Rv} HJD$-2400000$ &  phase &$V_{\rm r}$ (km s$^{-1}$)&Spectrograph \\ \hline
& 50835.6937&  0.7901&    45.80{\Rh}&     REOSC \\
& 50836.6782&  0.8363&    43.50{\Rh}&     REOSC \\
& 53663.8587&  0.6686&    46.38{\Rh}&     FEROS \\
& 53664.7578&  0.7109&    47.26{\Rh}&     FEROS \\
& 53665.7295&  0.7565&    47.57{\Rh}&     FEROS \\
& 53666.7902&  0.8064&    46.65{\Rh}&     FEROS \\
& 53890.4733&  0.3159&    20.34{\Rh}&     REOSC \\
& 53890.4882&  0.3166&    20.91{\Rh}&     REOSC \\
& 53891.4820&  0.3633&    26.92{\Rh}&     REOSC \\
& 53893.4307&  0.4549&    34.05{\Rh}&     EBASIM\\
& 53894.4388&  0.5022&    38.96{\Rh}&     EBASIM\\
& 54462.7415&  0.2034&     2.10{\Rh}&     REOSC \\
& 54682.9258&  0.5485&    38.32{\Rh}&     UVES  \\
& 54683.9155&  0.5950&    40.63{\Rh}&     UVES  \\
& 54814.8034&  0.7446&    45.48{\Rh}&     HARPS \\
& 54907.6066&  0.1049&   -13.63{\Rh}&     REOSC \\
& 54921.5702&  0.7610&    45.05{\Rh}&     HARPS \\
& 54985.4926&  0.7643&    44.72{\Rh}&     HARPS \\
& 54985.5149&  0.7654&    44.72{\Rh}&     HARPS \\ \hline
\end{tabular}
\end{center}
\end{table}

\begin{figure}
\includegraphics[width=54mm,height=84mm,angle=270]{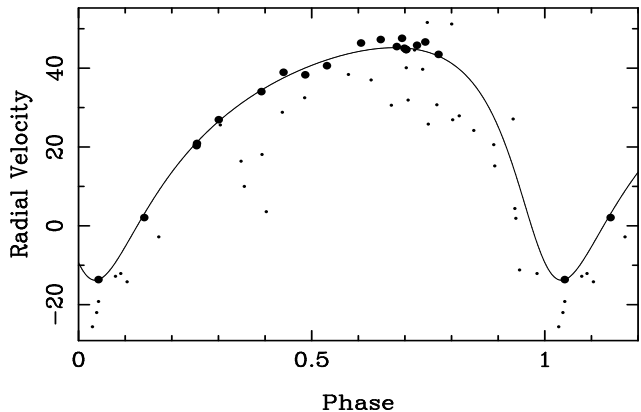}
 \caption{Old (dots) plus new (filled circles)
 radial velocities using orbital elements
 of Table~\ref{tab:sb1}.  A closer fit to the new data may
 be obtained if the period is determined only from the new
 data.  The current plot highlights the
 $\gamma$-velocity difference between the old and new data.
\label{fig:sb1}}
\end{figure}

HD 65949 was one of the objects investigated in
NGC 2516 for binarity by Abt and Levy (1972).
Their observations were combined with objective
prism measurements by Gieseking (1978) and
Gieseking and Karimie (1982),
who found orbital elements very close to those
adopted here (Table~\ref{tab:sb1}).  We may group
the radial velocities roughly into two time periods.
The ``old'' measurements were made within the interval
from Nov. 1967 through Apr. 1978.  ``New'' measurements
followed some 20 years later, from Jan. 1998 to
the Jun. 2009.  The newer measurements are clearly
more precise, as shown in Fig.~\ref{fig:sb1}.  This
is expected, as many of the older measurements were
made from objective prism spectra.

Table~\ref{tab:newrv} gives  the more recent,
previously unpublished, radial
velocities, plotted in Fig.~\ref{fig:sb1}.  The ESO FEROS
and UVES instruments are discussed in \S\ref{sec:spectra}.
The REOSC spectrograph is described by Gonz\'{a}lez and
Lapasset (2000); Pintado and Adelman (2003) discuss the EBASIM
instrument.

Even though the old measurements are less precise by a
factor 10--100, they
are useful for the period calculation since they provide
a time-base of about four decades.
We performed a global fit of all the observations to determine
the period.
Then we kept the period fixed and fit the remaining parameters
using only the new measurements.
The resulting parameters are listed in Table~\ref{tab:sb1}.

\begin{table}
\caption{\label{tab:sb1} SB1 orbital elements
and corresponding mass function, $f(m)$, for HD 65949.
Elements other than the period use only recent data.}

\begin{center}
\begin{tabular}{lrlc}\hline
Element  & Value &\multicolumn{2}{c}{Error} \\ \hline
$V_\gamma$ (km\,s$^{-1}${\Rh})&    25.7&{\Rh}$\pm$& 1.9 \\
$K_1$ (km\,s$^{-1}$)       &    29.5  &{\Rh}$\pm$& 1.4 \\
$\omega$ (deg)             &    148   &{\Rh}$\pm$& 7 \\
$e$                        &    0.40  &{\Rh}$\pm$& 0.05 \\
$P$ (d)                    &    21.2836&{\Rh}$\pm$&0.0012\\
$f(m)$  (M$_\odot$)       &    0.0437  &{\Rh}$\pm$&0.0067 \\
\hline
\end{tabular}
\end{center}
\end{table}

The adopted
$T_{\rm eff} = 13100$ K, and a fit to the data of Torres,
Andersen, and Gim\'{e}nez (2009) then yields
$M_V = -0.02$, for a main sequence star.  The
corresponding mass is
$\approx 3.3\,M_\odot$.
Since no absorption lines from the secondary are seen,
we assume a flux ratio $\le 0.1$, or $\Delta M_v > 2.5$.
The calibration gives $M/M_\odot \approx 1.52$, for
$M_V \approx 2.48$,
commensurate with a mass ratio, $q < 0.5$,
expected for main-sequence binary stars.   If we fix
the primary mass at 3.3 $M_\odot$,
the mass function, $f(m) = 0.0446$, yields a secondary mass
between 1.52 and 0.92 $M_\odot$ for inclinations in the
range 41 to 90$^\circ$.  Smaller inclinations are less
likely, since they would lead to larger secondary masses.
Masses of 3.3 $M_\odot$ for the primary, and from
1.52 to 0.92 $M_\odot$ for the secondary give separations
$a_1 + a_2$ from 0.254 to 0.243 AU.


The binarity of the HD 65949 system is of interest in view
of the suggestion made below that mass exchange may be
relevant for the surface chemistry.  We note the increase
in the systemic velocity of the binary system, which appears
convincing in spite of the larger scatter of the older
measurements.  Note also that there is a systematic
difference of 2 \kms between the FEROS and HARPS
observations, taken at the same phase, but separated by
3.5 years.

A fit of the center-of-mass velocity, keeping all other
orbital parameters fixed, gives
$V_\gamma=16.9\pm 1.5$ km\,s$^{-1}$ and
$V_\gamma=25.7\pm 0.2$ km\,s$^{-1}$ for the old and the new
measurements, respectively.  A third body is therefore
suspected to account for the change in systemic velocity.

\section{Atomic parameters}

Most of the atomic lines used in the present study
were sufficiently
weak that damping parameters are not important.  We
used default Stark damping from Cowley (1971), and Uns\"{o}ld's
(1955) formula for van der Waals damping, but enhanced by a
factor of two.  Only the Ca {\sc ii} K-line and lines from the infrared
triplet were strong enough that Stark damping began to be
relevant.  We used the parameters adopted by CH04.

Default oscillator strengths were from VALD (Kupka et al.
1999), but supplemented as noted in the element-by-element
discussion in Appendix~\ref{app:abs}.
Special calculations of partition functions and  oscillator
strengths were made for the present study, as noted in
the Appendices~\ref{app:pfn} and~\ref{app:loggf}.

For Cr {\sc ii}, Ti {\sc ii}, and Mn {\sc ii}, we used
VALD {\bf and Kurucz (1995) to
retain} only lines that were allowed by $LS$-coupling
selection rules.  All oscillator strengths for third spectra
of the lanthanides were from the DREAM database (Bi\'{e}mont,
Palmeri, and Quinet 1999).

\section{Abundances}

Table~\ref{tab:aabbs} lists abundances or upper
limits for 58 elements listed in Column 1.  Logarithmic ratios of
individual abundances to
the total elemental abundances including hydrogen follow
in Column 2.  Column 3 gives error estimates, which are
usually the standard deviation of the results from the
number of lines used, shown in Column 4.  For a few elements,
the error is the difference in determinations from two
ionization stages.  There is no entry for upper limits
based on a single line, but we estimate an uncertainty of
some 0.5 dex.  The solar abundance from Asplund, Grevesse,
Sauval and Scott (2009) is in Column 5, while Column 6 is
the difference in the stellar and solar values.

\begin{table}{\Rh}
\caption{\label{tab:aabbs}Abundances in HD 65949, the solar
system, and their differences (see text).  Results for
individual ions may be found in Appendix~\ref{app:abs}.}
\begin{tabular}{l  c  c  r c c} \hline
El{\Rv}  & $\log(N/\Sum)$ &$\pm$(sd)&n&$(\log(N/{\Sum})_\odot$   &[$N$] \\
\hline
He  &$    -1.95$&0.14& 8&$ -1.11 $&$    -0.84  $: \\   
C   &$    -3.67$&0.40& 2&$ -3.61 $&$    -0.06  $\\   
N   &$\le -6.52$&    & 1&$ -4.21 $&$    -2.31  $: \\  
O   &$    -3.24$&0.16& 9&$ -3.35 $&$    0.11   $:\\  
Ne  &$    -4.29$&0.16&10&$ -4.11 $&$    -0.18  $ \\  
Na  &$    -5.42$&0.13& 2&$ -5.80 $&$    0.38   $\\   
Mg  &$    -4.80$&0.45&13&$ -4.44 $&$    -0.36  $\\   
Al  &$    -6.44$&0.24& 4&$ -5.59 $&$    -0.85  $ \\ 
Si  &$    -4.69$&0.26&12&$ -4.53 $&$    -0.16  $\\   
P   &$    -5.13$&0.29&19&$ -6.63 $&$    1.50   $\\   
S   &$    -4.92$&0.24&34&$ -4.92 $&$    0.00   $\\   
Cl  &$\le -7.09$&    & 1&$ -6.54 $&$  \le  -0.55  $\\   
Ar  &$\le -6.07$&    & 1&$ -5.64 $&$  \le  -0.43  $: \\  
Ca  &$    -5.81$&0.53& 8&$ -5.70 $&$    -0.11  $:\\  
Sc  &$    -8.18$&0.07&  &$ -8.89 $&$    0.71   $\\   
Ti  &$    -6.89$&0.28&54&$ -7.09 $&$    0.20   $\\   
V   &$\le -8.65$&    & 1&$ -8.11 $&$\le -0.54  $\\ 
Cr  &$    -5.87$&0.31&63&$ -6.40 $&$    0.53   $ \\   
Mn  &$    -6.06$&0.15&22&$ -6.61 $&$    0.55   $ \\   
Fe  &$    -4.06$&0.24&98&$ -4.54 $&$    0.48   $\\   
Co  &$\le -5.76$&    &  &$ -7.05 $&$\le 1.29   $\\ 
Ni  &$    -6.35$&0.31& 5&$ -5.82 $&$    -0.53  $\\   
Cu  &$\le -5.81$&    & 1&$ -7.85 $&$\le 2.04   $\\ 
Zn  &$\le -7.8 $&    & 1&$ -7.48 $&$\le -0.32  $ \\
Ga  &$\le -7.50$&    & 1&$ -9.00 $&$\le 1.50   $\\ 
Br  &$    -6.81$&0.50& 3&$ -9.50 $&$    2.69   $\\   
Kr  &$    -5.85$&0.13& 5&$ -8.79 $&$    2.94   $\\   
Rb  &$\le -6.70$&    & 1&$ -9.52 $&$ \le 2.82  $ \\ 
Sr  &$    -6.76$&0.45& 6&$ -9.17 $&$    2.41   $ \\   
Y   &$    -7.66$&0.11&14&$ -9.83 $&$    2.17   $\\   
Zr  &$    -7.90$&0.17& 3&$ -9.46 $&$    1.56   $ \\  
Nb  &$    -7.26$&0.29&22&$ -10.58$&$    3.32   $\\   
Mo  &$    -7.86$&0.34& 4&$ -10.16$&$    2.30   $ \\  
Ru  &$    -6.50$&0.48&20&$ -10.29$&$    3.79   $\\   
Rh  &$    -7.43$&    & 1&$ -11.13$&$    3.70   $ \\ 
Pd  &$    -5.84$&0.14& 4&$ -10.47$&$    4.63   $\\   
Cd  &$\le -7.82$&    & 1&$ -10.33$&$ \le 2.51  $\\  
Sn  &$\le -8.42$&    & 1&$ -10.00$&$ \le 1.58  $\\
Xe  &$ {\bf -5.42}$&{\bf 0.11}& 6&$ -9.80 $&${\bf 4.38}   $\\   
Cs  &$    -7.50$&    & 1&$ -10.96$&$    3.46   $ \\
Ba  &$\le -9.64$&    & 1&$ -9.86 $&$    0.22   $ \\  
Ce  &$\le -9.79$&    & 1&$ -10.46$&$    0.67   $ \\  
Pr  &$    -8.31$&0.21&16&$ -11.32$&$    3.01   $\\   
Nd  &$    -8.03$&0.32&12&$ -10.62$&$    2.59   $\\   
Eu  &$\le -8.50$&    & 1&$ -11.52$&$\le 3.02   $\\
Dy  &$    -8.06$&0.44&12&$ -10.94$&$    2.88   $\\   
Ho  &$    -8.18$&0.31&12&$ -11.56$&$    3.38   $\\   
Er  &$    -8.80$&0.21& 3&$ -11.12$&$    2.32   $\\   
Yb  &$    -8.69$&    & 1&$ -11.20$&$    2.51   $\\   
W   &$\le -8.14$&    & 1&$ -11.19$&$\le 3.05   $ \\ 
Re  &$    -5.81$&0.27&32&$ -11.78$&$    5.97   $ \\   
Os  &$    -5.27$&0.53&13&$ -10.64$&$    5.37   $ \\  
Pt  &$    -5.22$&0.15& 6&$ -10.42$&$    5.20   $ \\  
Au  &$    -6.96$&0.52& 3&$ -11.12$&$    4.16   $\\   
Hg  &$    -4.59$&0.29& 4&$ -10.87$&$    6.28   $ \\  
Pb  &$\le -8.12$&    & 1&$ -10.29$&$\le 2.17   $ \\ 
Bi  &$\le -8.00$&0.50& 2&$ -11.39$&$\le 3.39   $\\
Th  &$    -9.18$&0.17& 8&$ -12.02$&$    2.84   $ \\
\hline
\end{tabular}
\end{table}

\begin{figure}
\includegraphics[width=54mm,height=84mm,angle=270]{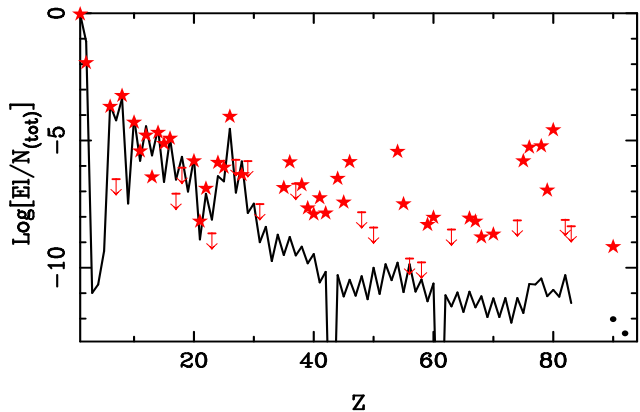}
 \caption{Solar (black line) and stellar (stars) abundances.
Upper limits are indicated by horizontal lines with arrows
pointing down.  Solar points for uranium and thorium are
filled circles.
\label{fig:abun}}
\end{figure}

The stellar abundances are plotted in Fig.~\ref{fig:abun},
along with corresponding solar values.  Generally small
deviations from the solar pattern are seen, especially for
the even-Z elements with Z less than about 30.
Beyond this point, the
stellar abundances scatter wildly, with excesses ranging
up to 6 dex (Re and Hg).

\section{Non-nuclear signatures}

The abundance pattern of Fig.~\ref{fig:abun} shows a number
of features that indicate the influence of non-nuclear
processes.  Interestingly, all such indications are for
elements with Z greater than about 30 (Zinc).  The most
common non-nuclear pattern shown in late-B CP stars is
an abundance of Mn (Z = 25), greater than that of either
Cr (Z = 24) or Fe (Z = 26).  This has been called an
{\it odd-Z anomaly}, since nuclear processes do not make
more of odd-Z elements relative their even-Z neighbors
(Li, Be, B excepted).  That anomaly at Mn is not seen in
HD 65949, but does appear in the typical HgMn star
HR 7143 (CH04).

Beyond the iron peak, there is often an odd-Z anomaly
at yttrium (Z = 39), which can be more abundant than its
even-Z neighbors, Sr and Zr (Guthrie 1971, Adelman et al.
2001).  This anomaly
is clearly present in HR 7143, but not in HD 65949.
However, the next triplet
containing an odd-Z element, Zr, Nb (Z = 41),
Mo does form an odd-Z
anomaly in HR 65949.
CH04 do not report
abundances for Nb and Mo.  Indeed, abundances for
these two elements are rarely (if ever) reported for HgMn
or related (HR 6000, HR 6870) stars.

Just as significant as the odd-Z anomalies are two highly
fractionated even-Z neighbors:
Xe and Ba.  We find
Xe more abundant than Ba by 4.2 or more
dex.  None of the standard (r- or s-process) neutron addition
schemes would produce so severe a fractionation.  Note that
a Xe-Ba fractionation of 3.31 dex occurs in HR 7143.
Similar (or larger) values probably hold for other HgMn stars
where Xe {\sc ii} has been identified.  However, the Ba {\sc ii} lines
are presumably not seen, and upper limits have not been
computed.

\section{Late B-star abundances compared}
\subsection{HD 65949 and HR 7143}

\begin{figure}
\includegraphics[width=54mm,height=84mm,angle=270]{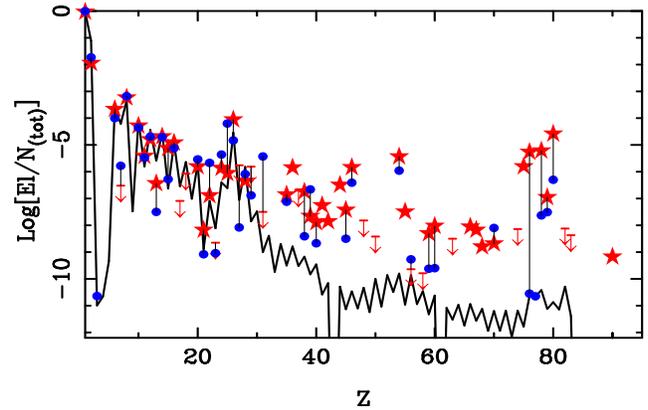}
 \caption{Solar abundances are again given by the solid black
line.  The stars or upper-limit symbols again indicate abundances
for HD 65949.  Filled circles show abundances for HR 7143.  When
elements are determined in both stars, a vertical line connects
the two values.  Note the enormous difference in the osmium
abundances (Z = 76).
\label{fig:both}}
\end{figure}

{\bf Figure}~\ref{fig:both} is a similar to Fig.~\ref{fig:abun}, but
shows abundances of both HD 65949 and HR 7143 (CH04).  Vertical
lines connect elements with abundances available for both stars.
Among the elements below zinc (Z = 30), the even-Z elements,
especially the lighter ones, are not far from their solar
values.  Larger departures from the solar pattern are seen
among the odd-Z elements.  These are usually small or negative
in HD 65949.
A nitrogen (Z = 7) deficiency is a general characteristic of HgMn
stars (Dworetsky 1993).
Phosphorus (Z = 15) is overabundant in both stars, but not more
so than its even-Z neighbors.  Manganese (Z = 25) is slightly overabundant
in HD 65949, but 2.5 dex in excess in HR 7143.
Overall, we may conclude that {\it the lighter elements
of HR 7143 show a greater fractionation from the solar pattern
than those of HD 65949}.

When we consider the heavier elements, both stars show a large
scatter, nearly exclusively of positive deviations from solar.
Gallium is particularly notable, as it is some 3.7 dex overabundant
in HR 7143.  The upper limit in HD 65949 is about 1.7 dex.
There is no indication of the stronger Ga {\sc ii} lines on the HARPS
spectra.  Apart from Ga and Y, the {\bf excesses} in HR 7143 are
lower than in HD 65949.  Note, especially the case for Os (Z = 76)
which is 5.3 dex in excess in HD 65949 but essentially solar
in HR 7143.

Generally, {\it among the elements heavier than zinc, the
abundances in HD 65949 are more highly fractionated than those
of HR 7143}.

\subsection{$\chi$ Lup}

Abundances for $\chi$ Lup A ($T_{\rm eff} = 10165$ K,
$\log(g) = 3.8$, B9.5p HgMn)
are given by Leckrone et al. (1999)
as supplemented by papers cited by Wahlgren (2005).
A plot of abundances vs. Z reveals a
relatively small scatter for most elements lighter than those
with Z in the early 30's.  This has been noted for HD 65949
and HR 7143.
There are, however, several
points for elements studied in the $HST$ UV spectra.
In particular, the marked {\it underabundance} of zinc (Z = 30)
is prominent.  The triplet Zn, Ga (Z=31), and Ge (Z = 32) form
an {\it even-Z} anomaly, incomprehensible from the point
of view of nucleosynthesis.  The same even-Z anomaly is
shown by Ga, Ge, and As.
No marked overabundances occur in $\chi$ Lup until Z = 33.
Beyond that value of Z, overabundances are common, and there
are no underabundances for detected elements.  {\bf Dworetsky,
Persaud \& Patel (2008) give $\log(Xe/H) = -5.74$, between
the values for HR 7143 and HD 65949.}
There {\bf is no abundance for the
noble gases Kr (Z=36)}.
Strontium (Z = 38) is enhanced in both HD 65949 and $\chi$ Lup.
Barium (Z=56) is significantly enhanced only in $\chi$ Lup
and in HgMn stars, but not in HD 65949.
Rhenium (Z = 75), so highly enhanced in HD 65949, has no
detection in $\chi$ Lup.  The overall very
heavy element (Os, Pt, Au, Hg, Tl) enhancement
{\bf in Chi Lup} is present,
but differs in detail from that of both HD 65949 and
HR 7143.

We leave further discussion and possible interpretation of the
$\chi$ Lup abundances to a future study.


\section{Discussion}
\subsection{The temperature differential}

We reject the temperature differential, some 1100 K,
as {\it primarily}
responsible for the abundance differences discussed in the previous
section.  That is because similar abundance patterns persist in
HgMn stars over comparable temperature ranges.
Moreover, strong Hg and
especially Pt are more common among cooler HgMn stars than
hotter ones, and these elements are more abundant in the hotter
star, HD 65949 than the cooler HR 7143.  A similar argument
applies to the Mn abundance, but with a reversed sense.  Here
the hotter HgMn stars are generally richer in Mn, but the cooler
HR 7143 has the larger Mn abundance excess.

The isotopic composition of Hg in HD 65949,
is also more typical of cooler HgMn stars than that of HR 7143.
At low resolution, a
mean wavelength of the Hg {\sc ii} feature may indicate an enrichment
of the heavier isotopes--generally the cooler
HgMn stars have longer center-of-gravity
wavelengths for the $\lambda$3984 feature.
However, for HR 7143, we measured a
mean position of 3983.858~\AA\, on a 2.4 \AA/mm plate taken
at the Dominion Astrophysical Observatory (9682/10858u).
This might be compared with the FEROS wavelength (cf. Paper I)
of 3984.01~\AA\, for HD 65949.  Synthesis of the higher-resolution
HARPS and UVES spectra
available today prove that HD 65949 is
richer in heavier Hg isotopes, though $^{204}$Hg does not
dominate, as in $\chi$ Lup.

\subsection{Nuclear patterns}

The elements Sr and Ba are typically associated with the
s-process.  While the Sr excess is more than 2 dex, Ba
is at most marginally enhanced, and could be depleted.
This excludes the relevance of that process.  On the
other hand, the solar system r-process shows excesses at
Te and Xe, and again, at Os and Pt.  The former peak is
associated with the N=82 neutron shell closing, and the latter
closed shell at
N=126.  We have not reported an abundance for Te, but the
element is positively identified, and surely in excess.
Oscillator strength calculations currently under way will
provide a quantitative result.

We have noted that the idea of mass transfer in connection
with CP star anomalies
is relatively old.  Wahlgren et al. (1995) discuss it briefly
in connection with the isotopic anomalies in $\chi$ Lup that
suggest the r-process.  We note also, the shrewd observation
of Woolf and Lambert (1999) that the stable, lighter Hg isotopes
are never enhanced in HgMn stars, and that these are the only
two isotopes {\it not} produced by the r-process.  On the
other hand,
Proffitt and Michaud (1989) concluded the likely transferrence
of a significant amount of material from a nearby supernova to
a B or A star was ``one in a few thousand.''
Even if HD 65949 represents that rare star, the abundances
of the heavy elements are severely fractionated
from a pure nuclear pattern.  {\it The anomalies cannot
result only from an admixture of nuclear-processed material.}

We therefore look within this pattern for some clue to the
relevant fractionation mechanism.  The favored mechanism
would be {\it in situ} separation by radiative and gravitationally
driven diffusion.

\subsection{Theoretical predictions}

Surprisingly little theoretical work is of relevance to the
present task.  An exception is the decades-old, but extensive
work by Michaud, Charland, Vauclair, \& Vauclair (1976, MCVV).
In this work both time scales, and extensive predictions
are made through the lanthanide elements.  In Fig.~6 of MCVV
there are precipitous abundance drops near Z = 38-40, and 56-58.
This suggests a depletion at Sr, which we do not see, and
one at Ba, which we may.
On the other hand, the
calculations for the heavier elements were sufficiently
rough that most elements were predicted to be overabundant
by about the same amount.  Only when relevant ions
achieved
the noble gas configurations (e.g. Sr {\sc iii} or Ba {\sc iii})
was there a significant reduction of the
radiative to gravitational
(plus temperature-gradient) forces
($g_R/g_{GT}$).  That ratio was more or less constant for
most of the elements beyond Z = 30, and could not account
for the structure seen in our Fig.~\ref{fig:abun}.

The basic diffusion hypothesis has always been that the
stars arrive on the main sequence with abundances that are
well mixed.  Chemical separation then
took place as a result of a time-dependent process.
Nevertheless, there has been almost no attempt to interpret
abundances in terms of age.

The concept of age, as we need it here,
need not be a chronological age, or years on the main sequence.
MCVV and many subsequent
studies (cf. Richer, Michaud, \& Turcotte 2000)
added a ``turbulent'' component
to the diffusion coefficient.  Without this modification
much larger anomalies than those observed in some CP
stars would be predicted.  The more effective
this turbulence, the slower the diffusion processes would be.
Thus we must think of age in a relative sense.  Chemical or
cosmochemical
maturity might be a more appropriate phrase than age.

For stars with masses above 2-2.6 $M_\odot$, MCVV found
characteristic diffusion times very short with respect to the
stellar lifetimes. They proceeded to predict abundance
anomalies for these stars, without consideration of
the ages or the length of time since the diffusion began
to operate.  Presumably, this was because the time scales
were found to be very short, after which time an approximate
equilibrium abundance pattern might be established.

This basic picture
does not account for the occurrence
of very different abundances in stars with similar temperatures
and surface gravities.

\subsection{Correlations}

Suppose exotic material from a supernova were transferred
to the surface of a nearby star.  We must imagine the
material to have characteristic r-process enhancements,
and minor amounts of elements with Z $<$ 30.

This material could be subject to grain condensation, followed
by gas-grain separation as in the scenario proposed by
Venn and Lambert (1990) to explain $\lambda$ Boo stars.
In our case, the elements that resist grain formation
(Xe, Kr, Os, Pt, Hg) would be carried to the star.
One might then expect to see a correlation of the abundance
excesses with condensation temperatures (e.g. Lodders, 2003).
However, we find no significant correlation of this kind.

We therefore favor a second model.



\begin{figure}
\includegraphics[width=54mm,height=84mm,angle=270]{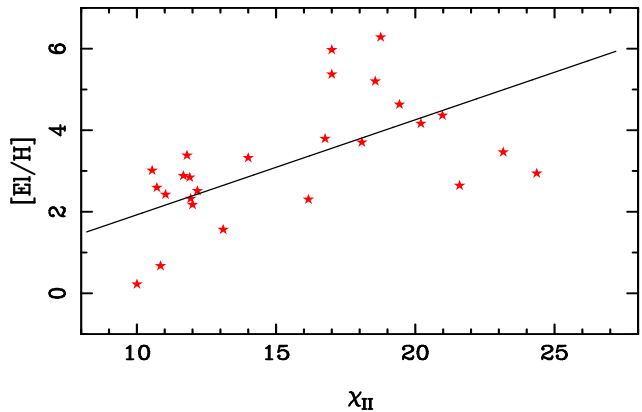}
 \caption{Logarithmic abundance excesses for elements with
Z $>$ 30 vs. second ionization energies.  Upper limits are
not included.  The line is a least squares fit.  The
correlation coefficient, 0.608, is significant at the 0.0013
level for 25 points.\label{fig:2ipcor}}
\end{figure}

{\bf Figure}~\ref{fig:2ipcor} shows that there is a correlation of the
abundance excesses of the heavier elements with the second
ionization potential.
The significance of the correlation
is 0.0013.  The figure and significance calculation includes
two points for Ba and Ce for which we have only upper limits.
Should we exclude them, the significance drops to 0.016.  However,
if we use abundances for these elements decreased by 1 and
even 2 dex, the
significance of the correlation is essentially the same as
with the upper limits: 0.0016.

While it is clear that other factors than the second ionization
potential determine the overabundances, the correlation we
find is unlikely to have arisen by chance.

The proposed model supposes that
exotic r-processed material fell onto the star, and {\it then}
was subject to in situ differentiation.  One advantage of this
model is that it would not require the diffusion of rhenium
or mercury from great depths requiring the elements to pass
through many ionization stages.

The ``mass above the photosphere'' may be defined as
$\int \rho dx$ from $x = 0$ to a physical depth where the
optical depth is about unity.  For late B stars, that mass
is about 0.1\, gm\,cm$^{-3}$.  This means that to enhance the
Hg abundance by a factor of $2\cdot10^6$, ions would have
to diffuse upward from a depth such that
$\int \rho dx \approx 2\cdot 10^5$.  We use
the 2.5 $M_\odot$ model of P. Demarque, D. Guenther,
and J. Howard (cf. Cowley 1995, Table 9.3(b)).  Numerical
integration of the tabulated values show that at this
depth ($r/R = 0.92$), $T = 4\cdot 10^5$K.  If we use
the Saha equation, taking the ratio of the relevant
partition functions to be unity, we find the ratio
of Hg$^{+16}$ to Hg$^{+15}$ approximately unity at this
depth ($9.6\cdot 10^4$\, km below the surface of the
star).

This calculation assumed mercury that has diffused upward
does not escape from the photosphere.  We also have neglected
lowering of the ionization energy.  Both effects would increase
the relevant degree of ionization.
The ionization
energy for Hg {\sc xvi} (+15) used here is from Carlson et al. (1970).
The value, 357 eV, is approximate, but adequate for the
present purposes, which is only to show the multiplicity
of ion stages involved.  The overall situation is not
substantially different from that discussed by
Cowley and Day (1976), where it was concluded that for an
enhancement of $10^5$ mercury would have to diffuse from
depths involving Hg {\sc xi}-{\sc xiii}, and the relevant temperature
(see Table 1. Model B) was $2\cdot 10^5$.

Diffusion from deep layers does not provide a basis for
understanding the correlation with the second ionization potential.

\subsection{Predictions}

Based on the model of the preceeding section, and
the abundances of HD 65949 and
HR 7143 (as typical of HgMn stars), we can make a number of
predictions that can be checked by further investigation.
The overall hypothesis is that both stars have been subject
to exotic mass addition, but that HD 7143 is cosmochemically
more mature then HD 65949.

\begin{itemize}
\item Osmium and rhenium are rarely (if ever) enhanced
to the point where they are identified in ground-based spectra
of HgMn stars.
Therefore, these elements cannot be (strongly)
supported by radiation in
the atmospheres of HgMn stars compared, for example, to mercury.
(Their high abundance in
HR 65949 must result from a very recent transfer of material.)

\item Xenon is often found in HgMn stars.  In spite of the
fact that Xe {\sc  i} has noble gas structure, significant support
for this element must exist.  The case of Kr needs more
observational material.  Though it is not seen in HgMn stars,
better observations could lead to its identification.  We
see no indication of Kr {\sc ii} $\lambda$4355.48 in HR 7143 on CH04 material
posted on Castelli's web site.

\item The nitrogen deficiency and the phosphorus excess must
be established rapidly in HD 65949.  These anomalies appear
before a significant Mn excess appears.

\item The large gallium abundance shown by many HgMn stars is
not seen in HD 65949.  Therefore, it must be pushed up from
considerable depths, on a time scale longer than relevant
for HD 65949.  It would be a pure diffusion anomaly.
\end{itemize}

Of course, it may be possible to account for the abundance
pattern of HR 65949 with the fundamental diffusion picture
when the necessary atomic data are known.
The lack of {\bf barium enhancement} may already
be explained in MCVV.  The overall scatter of the heavier
elements could be attributed to the relative ease of
transport of atoms with initially low abundances.
This would
mean that mercury and rhenium were pushed up from deep layers
where atoms might be ten or more fold ionized.  We would
then conclude that the observed correlation with second
ionization potential was a red herring.

\section{Acknowledgements}

We thank P. North and B. Smalley for computer codes,
and J. R. Fuhr J. Reader, and W. Wiese of NIST for advice on
atomic data and processes.  Thanks also to J. J. Cowan
for comments on the abundance pattern in HD 65949
from the point of view of nucleosynthesis.
This research has made use of the SIMBAD database, operated
at CDS, Strasbourg, France.
Our calculations have made extensive use of
the VALD atomic data base (Kupka et al. 1999).    Thanks
are also due to M. Netopil for help during observations
in August of 2008.  CRC thanks John Hutchings,
Murray Fletcher, and his
colleagues at Michigan, especially Ming Jhao,
for numerous ideas and comments.  Thanks are due to J.
Andersen and G. Torres for a preprint of their review paper
on the masses and radii of normal stars.  Financial
support of the Belgian FRS-FNRS is also acknowledged.
GMW acknowledges support from NASA grant NNG06GJ29G.

\section*{REFERENCES}

\hangpar Abt, H. A., Levy, S. G. 1972, ApJ, 172, 355

\hangpar Abt, H. A., Morgan, W. W. 1969, AJ, 74, 813

\hangpar Adelman, S. J., Snow, T. P., Wood, E. L., Ivans, I. I.,
Sneden, C., Ehrenfreund, P., Foing, B. H. 2001, MNRAS, 328, 1144


\hangpar Asplund, M., Grevesse, N., Sauval, A. J., Scott, P.
2009, ARA\&A. 47, 481

\hangpar Blaise, J., Wyart, J.-F. 2009, online data tables:
\newline http://www.lac.u-psud.fr/Database/Contents.html

\hangpar Bi\'{e}mont, E., Grevesse, N., Kwiatowski, M.,
Zimmerman, P. 1982, A\&A, 108, 127


\hangpar Bi\'{e}mont, E. Palmeri, P., Quinet, P. 1999,
Astrophys. Sp. Sci., 269-270, 635.  See also the online
data base: (http://w3.umh.ac.be/$\sim$astro/dream.shtml)



\hangpar Carlson, T. A., Nestor, C. W. Jr., Wasserman, N.
McDowell, J. D. 1970, Atomic Data, 2, 63

\hangpar Castelli, F. 2009, wwwuser.oats.ts.astro.it/castelli/

\hangpar Castelli, F., Hubrig, S. 2004, A\&A, 425, 263 (CH04)

\hangpar Cowan, R. D. 1981, The Theory of Atomic Structure
and Spectra (Berkeley, CA: Univ. Calif. Press)

\hangpar Cowley, C. R. 1971, Obs., 91, 139

\hangpar Cowley, C. R. 1995, An Introduction to Cosmochemistry,
(Cambridge: University Press)

\hangpar Cowley, C. R., Barisciano, L. P., Jr. 1994, Obs.,
114, 308

\hangpar Cowley, C. R., Day, C. A. 1976, ApJ, 205, 440

\hangpar Cowley, C. R., Wahlgren, G. M. 2006, A\&A, 447, 681



\hangpar Cowley, C. R., Hubrig, S., Wahlgren, G. M. 2008,
J. Phys. Conf. Ser., 130, 12005 (Paper II)

\hangpar Cowley, C. R., Hubrig, S., Gonz\'{a}lez, G. F.,
Nu\~{n}ez, N. 2006, A\&A, 455, L21 (Paper I)

\hangpar Dekker, H., et al. 2000, SPIE, 4008, 534

\hangpar Dolk, L., Litz\'{e}n, U., Wahlgren, G. M. 2002,
A\&A, 388, 692

\hangpar Dworetsky, M. M. 1993, in Peculiar Versus Normal
Phenomena In A-Type And Related Stars, ed. M. M. Dworetsky,
F. Castelli, and R. Faraggiana, ASP Conf. Ser., 44, 1

\hangpar Dworetsky, M. M., Storey, P. J., Jacobs, J. M. 1984,
Phys. Scr. T8, 39

\hangpar Dworetsky, M. M., Persaud, J. L., Patel, K. 2008,
MNRAS, 385, 1523

\hangpar Engleman, R. 1989, ApJ, 340, 1140



\hangpar Fivet, V., Quinet, P., Bi\'emont, \'E., Xu, H. L.
2007, J. Electron. Spec. Relat. Phen., 156-158, 250

\hangpar Fuhr J. R., Wiese, W. L. 1996, NIST Atomic Transition
Probability Tables, CRC Handbook of Chemistry \& Physics,
77th Edition, ed. D. R. Lide, CRC Press, Inc., Boca Raton, FL

\hangpar Fuhr, J. R., Wiese, W. 2006, J. Phys. Chem. Ref. Data,
35, 1669

\hangpar Gieseking, F. 1978, A\&AS, 32, 17

\hangpar Gieseking, F., Karimie, M. T. 1982, A\&AS, 49, 497



\hangpar Gonz\'{a}lez, J. F., Lapasset, E. 2000, AJ, 119, 2296

\hangpar Grevesse, N. 2008, in {\it Comm. in Astroseismology},
Vol. 157, 156 (Wrocklaw HELAS Workshop), ed. M. Breger,
W. Dziembowski, \& M. Thompson.

\hangpar Guthrie, B. N. G. 1971, Astrophys. Sp. Sci., 10, 156

\hangpar {\bf Hauck, B., Mermilliod, M. 1998, A\&AS, 129, 431}


\hangpar Hubrig, S., North, P., Sch\"{o}ller, M.,
Mathys, G. 2006, AN, 327, 289

\hangpar Ivarsson, S., Wahlgren, G. M., Dai, Z., Lundberg, H.,
Leckrone, D. S. 2004, A\&A, 425, 353


\hangpar Kaufer, A., {\bf Stahl, O., Tubbesing, S.,
N{\o}rregaard, P., Avila, G., Francois, P.,
Pasquini, L., Pizzella, A.}  1999, ESO Messenger, 95, 8

\hangpar Klinkenberg, P.F.A., Meggers, W.F., Velasco, R.,
Catalan, M.A. 1957,  J. Res. Natl. Bur. Stand., 59, 319

\hangpar Kramida, A. E., Shirai, T. 2006, J. Phys. Chem. Ref.
Data, 35, 423

\hangpar Kunzli, M., North, P., Kurucz, R. L., Nicolet, B. 1997,
A\&AS, 122, 51

\hangpar Kupka, F., Piskunov, N. E., Ryabchikova, T. A.,
Stempels, H. C., Weiss, W. W. 1999, A\&AS, 138, 119

\hangpar Kurucz, R. L. 1993, CD-ROM No. 18 (Smithsonian
Ap. Obs.)

\hangpar Kurucz, R. L. 1995, CD-ROM No. 23 (Smithsonian
Ap. Obs.)

\hangpar Leckrone, D. S., Proffitt, C. R., Wahlgren, G. M.
Johansson, S. G., Brage, T. 1999, AJ., 117, 1454




\hangpar Ljung, G., Nilsson, H., Asplund, M., Johansson, S.
2006, A\&A, 456, 1181

\hangpar Lodders, K. 2003, ApJ, 591, 1220

\hangpar Mayor, M., {\bf Pepe, F., Queloz, D., et al.} 2003,
ESO Messenger, 114, 20

\hangpar Meggers, W. F., Catalan, M. A., Sales, M. 1958,
J. Res. Natl. Bur. Stand., 61, 441

\hangpar Meggers, W. F., Corliss, C. H., Scribner, B. F.
1975, NBS Monog. 145

\hangpar Mel\'{e}ndez, J., Barbuy, B. 2009, A\&A, 497, 611


\hangpar Mermilliod, J.-C., Hauck, B., Mermilliod, M. 2007,
General Catalogue of Photometric Data,
\noindent http://www.unige.ch/siences/astro/

\hangpar Michaud, G. 1970, ApJ, 160, 641

\hangpar Michaud, G., Charland, Y., Vauclair, S., Vauclair, G.
1976, ApJ, 210, 447 (MCVV)

\hangpar Moon, T. T. 1984, Comm. Univ. London Obs., No. 78

\hangpar Moon, T. T., Dworetsky, M. M. 1985, MNRAS, 217, 305

\hangpar Moore, C. E. 1949-1958, Atomic Energy Levels, Vols. I-III,
NBS Circ. 467 (Washington, D. C.: US Gov. Print. Off.)

\hangpar Munari, U., Zwitter, T. 1997, A\&A, 318, 269

\hangpar Nilsson, H., Ivarsson, S. 2008, A\&A, 492, 609

\hangpar Nilsson, H., Ljung, G., Lundberg, H., Nielsen,
K. E. 2006, A\&A, 445, 1165

\hangpar Palmeri, P., Quinet, P., Bi\'emont, \'E.,
Svanberg, S., Xu, H.L. 2006, Phys. Scr., 74, 297

\hangpar Palmeri, P., Quinet, P., Bi\'emont, \'E., Xu, H.L.,
Svanberg, S. 2005, MNRAS, 362, 1348

\hangpar Palmeri, P., Quinet, P., Fivet, V.,
Bi\'{e}mont, \'{E}., Cowley, C. R., Engstr\"{o}m, L.,
Lundberg, H., Hartman, H, Nilsson, H. 2009, J. Phys
B 42, 165005

\hangpar Paunzen, E., Schnell, A., Maitzen, H. M. 2006, A\&A,
458, 293

\hangpar Pickering, J. C., Thorne, A. P., Perez, R. 2001,
ApJS, 132, 403 (Erratum: ApJS, 138, 247, 2002)

\hangpar Pintado, O. I., Adelman, S. J. 2003, A\&A, 406, 987

\hangpar Preston, G. W. 1974, ARA\&A, 12, 257

\hangpar Proffitt, C. R., Michaud, G. 1989, ApJ, 345, 998

\hangpar Quinet, P., 2002, J. Phys. B., 35, 19

\hangpar Quinet, P., Palmeri, P., Bi\'emont, \'E., Jorissen,
A., Van Eck, S., Svanberg, S., Xu, H.L., Plez, B. 2006,
A\&A, 448, 1207

\hangpar Quinet, P., Palmeri, P., Fivet, V., Bi\'{e}mont, \'{E}.,
Nilsson, H., Engstr\"{o}m, L, Lundberg, H. 2008, Phys. Rev. A.,
77, 022501


\hangpar Ralchenko, Yu., Kramida, A.E., Reader, J.
    and NIST ASD Team 2008, NIST Atomic Spectra Database
    (version 3.1.5), [Online]. Available:
    http://physics.nist.gov/asd3 [2010, Jan 24]. National
    Institute of Standards and Technology, Gaithersburg, MD.

\hangpar Reader, J., Corliss, C. H. 1980, NSRDS-NBS Monog. 68.

\hangpar Richer, J., Michaud, G., Turcotte, S. 2000, ApJ, 529,
338

\hangpar Rosberg, M., Wyart, J.-F. 1997, Phys. Scr., 55, 690.




\hangpar Ryabtsev, A.N. 2009, (private communication)

\hangpar Ryabchikova, T., Ryabtsev, A., Kochukhov, O.,
Bagnulo, S. 2006, A\&A,
456, 329

\hangpar Sansonetti, C. J., Reader, J. 2001, Phys. Scr., 63, 219

\hangpar Smirnov, Yu. M., Shapochkin, M. B. 1979, Opt. Spectroc.,
47, 243


\hangpar Torres, G., Andersen, J., Gim\'{e}nez, 2010,
A\&A Rev., 18, 67

\hangpar Uns\"{o}ld, A., 1955, Physik der Sternatmosph\"{a}ren,
Zweite Aufl. (Berlin: Springer)

\hangpar van Leeuwen, F. 2009, A\&A, 497, 209

\hangpar Venn, K. A., Lambert, D. L. 1990, ApJ, 363, 234

\hangpar Wahlgren, G. M. 2005, in The A-Star Puzzle, ed.
J. Zverko, J. \v{Z}i\v{z}\v{n}ovsk\'{y}, S. J. Adelman,
W. W. Weiss (Cambridge: Cambridge Univ. Press.), p. 291

\hangpar Wahlgren, G. M., Johansson, S., Litz\'{e}n, U.,
Gibson, N. D., Cooper, J. C., Lawler, J. E., Leckrone, D.,
Engleman, R. Jr. 1997, ApJ, 475, 380

\hangpar Wahlgren, G. M., Leckrone, D. S., Johansson, S. G.,
Rosberg, M., Brage, T. 1995, ApJ, 444, 438

\hangpar Wolk, S. J., Harnden, F. R. Jr., Murray, S. S.,
et. al. 2004, ApJ, 606, 466

\hangpar Woolf, V., Lambert, D. L. 1999, ApJ, 521, 414

\hangpar Wyart, J.-F. 1977, Optica Pura y Aplicada {\bf 10}, 177

\hangpar {\bf Zieli\'{n}ska, S., Bratasz, {\L}.,
Dzier\.{z}\c{e}ga, K. 2002, Phys. Scr., 66, 454}


\appendix
\section{Discussion of individual abundances}
\label{app:abs}

The Michigan software used to obtain abundances from
individual lines is set up to read a data base that
is basically VALD, but with numerous additions and
edits.  For example, for the third spectra of the
lanthanides, all values are from the DREAM site.  Use
of data from original sources often requires line-by-line
editing.  To avoid this in many cases, we used our
default data base, but checked against more recent
sources, or against the posting on the NIST site, and
if the differences were minor, we did not recompute
an abundance.

All averages are logarithmic, that is, the logarithms of
abundances for individual lines were averaged directly.

Original sources of oscillator strengths are cited
where practicable, but we relied heavily on the
online data bases of Ralchenko, et al. (2008,
NIST), Kupka, et al.
(1999, VALD), and Bi\'{e}mont, Palmeri \& Quinet
(1999, DREAM).  When no source of oscillator strength
is explicitly cited, the values come from VALD.

In the following sections, a measured stellar wavelength
is indicated by an asterisk, e.g. $\lambda^*$4911.66.

{\bf Helium (Z = 2; $\log(He/\Sum) \approx -1.95 \pm 0.14$):}
A rough estimate, using Voigt profiles of 8 He {\sc i} lines.
The helium abundance is about 10 per cent that of the sun.

{\bf Carbon (Z = 6; $\log(C/\Sum) = -3.66\pm 0.4$:):} The
abundance is from a synthesis of the  C {\sc ii} $\lambda$4267
doublet, which is clearly present.  The VALD oscillator
strengths are very close to those on the NIST site.
A few C {\sc i} and {\sc ii}
lines are surely present, but give inconsistent abundances,
most likely due to blends.
The carbon abundance
is nearly solar.  CH04 find a ca. 0.5 dex deficiency of
carbon.

{\bf Nitrogen (Z = 7; $\log(N/\Sum) \le -6.52$):}
Neither N {\sc i} nor N {\sc ii} can be positively
identified.  An upper limit of 0.3 m\AA\, for
N {\sc ii} $\lambda$8680 gives $\log(N/N_{\rm tot}) = -6.52$,
corresponding to a deficiency of about 2.4 dex
with respect to the sun. CH04 got an upper limit
corresponding to a deficiency of $-$1.7 dex in
HR 7143.

{\bf Oxygen (Z = 8; $\log(O/\Sum) = -3.24\pm 0.16$):} Nine
O {\sc i} lines, excluding the strong triplet
$\lambda\lambda$7772, 7774, and 7775 yield a small oxygen
excess above to the solar value.  Oscillator strengths
are from VALD but agree well with NIST.
CH04 also find
a slight excess of oxygen for HR 7143.

{\bf Neon (Z = 10; $\log(Ne/\Sum) = -4.29\pm 0.16$):}
A close examination of the HARPS spectra show that Ne I
is clearly present.  The abundance is based on 10
weak lines and oscillator strengths from NIST.  The
line-to-line agreement is excellent.

{\bf Sodium (Z = 11; $\log(Na/\Sum) = -5.42\pm 0.13$):}  The
abundance is from the  D-lines, with equivalent widths
of 27 and 21 m\AA. The probable error is the difference
of the two determinations.
The stellar lines are much weaker than the interstellar
ones.

{\bf Magnesium (Z = 12; $\log(Mg/\Sum) = -4.80\pm 0.45$):}
The adopted abundance is primarily from   9
Mg {\sc ii} lines, which give $-4.91\pm 0.17$.  We did not
use the $\lambda$4481 doublet for
an abundance.
Four Mg {\sc i} lines, including the b-lines give
$-4.46\pm 0.47$, which might indicate stratification
or too hot a model.
They are weighted 1/3 in the adopted
value.  The adopted error is the difference in the Mg I
and Mg {\sc ii} values.  Oscillator strengths are from VALD,
but are very close to those at the NIST site.

{\bf Aluminum (Z = 13; $\log(Al/\Sum) = -6.45\pm 0.24$):}
The two strongest Al {\sc ii} lines,
$\lambda\lambda$4663.05 and 6243.36 were measured and so
identified in the online wavelength list: \linebreak
\mbox{http:\\www.astro.lsa.umich.edu/$\sim$cowley/hd65949/}
\newline A 5.8 m\AA\,
line is present at the position of the strong Al I
line $\lambda$3691.52.  This is almost certainly the
line 3691.49 on the online list.  There is no
definite feature at the position of the other strong
Al {\sc i} line, $\lambda$3944.01.  However, the noise might
obscure a 1 m\AA\, feature.  We adopt a straight average
of two Al {\sc i} and two Al {\sc ii} lines.
Oscillator
strengths were from VALD, but are very close to NIST.
Aluminum is deficient.  This result was also found by CH05.

{\bf Silicon (Z = 14; $\log(Si/\Sum) = -4.69\pm 0.26$):}
The abundance is from 10 Si {\sc ii} and 2 Si {\sc iii}
lines with equivalent widths ranging from 4.6 to
136 m\AA.    Oscillator strengths are mostly from
NLTELINES (Kurucz 1993).
Generally, the agreement with NIST
was good.  However for $\lambda\lambda$5669
and 5957, we substituted the NIST values, which
had B+ accuracy.

The 2 Si {\sc iii} lines, $\lambda\lambda$4552
and 4567 yield abundances of $-$4.29 and $-$4.46,
in fair agreement with the overall mean.
Both lines have NIST graded B+ accuracies.

{\bf Phosphorus (Z = 15; $\log(P/\Sum) = -5.13\pm 0.29$):} The
abundance is based on 19 weak P {\sc ii} lines.
Oscillator
strengths are from BELLLIGHT
(Kurucz 1993), a compilation from various sources.
The overabundance, some 1.6 dex, is
substantially above that found by CH04 for
HR 7143.

{\bf Sulfur (Z = 16; $\log(S/\Sum) = -4.92\pm 0.24$):}  The
abundance is based on 34 S {\sc ii} lines with
equivalent widths from 1.3 to 17 m\AA.  Two outliers,
$\lambda\lambda$4162 and 4174 were excluded from the
average.
The abundance
is solar, within the uncertainties.
CH04 find
sulfur underabundant by 0.4 dex in HR 7143.

{\bf Chlorine (Z = 17; $\log(Cl/\Sum) \le -7.09$):}
An upper limit is derived from Cl {\sc ii}
$\lambda$4794.55, which is $\le 0.3$ m\AA.  The
oscillator strength is from Fuhr and Wiese (1996).

{\bf Argon (Z = 18; $\log(Ar/\Sum) \le -6.07$):} An upper
limit is derived from the strong
Ar {\sc i} line $\lambda$8115.31.  The oscillator strength
is from Fuhr and Wiese (1996).

{\bf Calcium (Z = 20; $\log(Ca/\Sum) = -5.81\pm 0.53$):}
The strongest Ca {\sc i} line, $\lambda$4227 is
probably present.  We measured a 1.6 m\AA\, line just
above the noise level.  There were {\bf seven} unblended Ca {\sc ii}
lines.  With a microturbulence, $\xi_t = 1$ \kms, the K-line and
$\lambda\lambda$8498 and 8542 of the infrared triplet
yield an abundance ca. 1 dex higher than the
weaker lines, $\lambda\lambda$3706, 8201, 8248,
and 8912.   The plot of abundance
vs. equivalent width looks like a classic case of
too low a microturbulence.  The adopted mean is the
average of assuming $\xi_t = 1$ and $\xi_t = 5$ \kms.  The
latter brings strong and weak Ca {\sc ii} lines into agreement.
The uncertainty is the difference in these abundances.
We do not consider $\xi_t = 5$ \kms realistic.
Interestingly, the
strong Ca {\sc ii} lines agree with the very weak Ca I
$\lambda$4227 line.  Calcium is poorly determined;
the source of the large uncertainty is
not understood.  The usual culprits are non-LTE or
stratification.  They are not explored here.
Note that the Ca {\sc ii} K-line and the two components
of the infrared triplet are the strongest lines used
in the present analysis.

{\bf Scandium (Z = 21; $\log(Sc/\Sum) = -8.18\pm 0.07$):}
The abundance is based on 5 Sc {\sc ii}
lines, including $\lambda$4246 with equivalent
widths from 5.1 to 16.8 m\AA.  The internal agreement
is good; the standard deviation for the {\bf five} lines
is less than 0.1 dex.

{\bf Titanium (Z = 22; $\log(Ti/{\Sum})=-6.90\pm 0.27$):}
Only Ti {\sc ii} lines are available.  The
abundance is based on 42 lines with equivalent
widths ranging from 2 to 45 m\AA.  Oscillator
strengths are from  Pickering, Thorne, \& Perez (2001, PTP).
Results were very similar if lines with
LS-allowed transitions from Kurucz (1995, used by VALD)
were used.  No
significant trends of abundances with wavelength,
equivalent width, or excitation potential were
noticed.  Three obvious outliers were excluded.  If
they are averaged in, the abundance would be $-$6.78.

{\bf Vanadium (Z = 23; $\log(V/\Sum)\le -8.65$):}
While V {\sc ii} lines are prominent in the spectra of
cooler Am and superficially normal A-stars; the lines are
typically weak or absent in hotter CP types.  We
estimate an upper limit taking the equivalent width
of V {\sc ii} $\lambda$3545.19 to be $\le 0.5$ m\AA.
This is close to the upper limit found for vanadium
by CH04.

{\bf Chromium (Z = 24; $\log(Cr/\Sum) = -5.87\pm 0.31$):}
There is
an $\approx$ 1 m\AA\, feature at the proper position
to be $\lambda$4254.35, the strongest Cr {\sc i} line in
the region.  That line alone gives an abundance of
$\log(Cr/N_{\rm tot}) = -5.6$ in satisfactory
agreement with
the overall mean of the Cr {\sc ii} lines.
The abundance
is based on 64 $LS$-permitted transitions.
Only 14 of the 64 lines used were found in the online
material published by Nilsson, et al. (2006).
A comparison of the {\it LS-permitted} lines in VALD
and Nilsson et al. (2006) gave a mean of +0.045 for
$\log(gf_{\rm VALD}) -\log(gf_{\rm Nilsson})$,
with a standard deviation of 0.16 dex.  This difference
was considered negligible at the present level of
accuracy.  There is a slight trend of abundance with
equivalent width that can be removed by assuming
$\xi_t=3 $\kms.  The resulting average abundance would
be $-5.98\pm 0.26$.  Since Cr has an even Z, and
hyperfine broadening is not anticipated, we retained
the result with $\xi_t = 1$ for consistency with other
spectra.

{\bf Manganese (Z = 22; $\log(Mn/\Sum) = -6.06\pm 0.15$):}
There is no indication of Mn I.
The abundance is based on 22 Mn {\sc ii} lines with
strengths ranging from 3.1 to 86 m\AA.  There was
no trend of abundance with equivalent width with
$\xi_t = 2.0$~\kms.  Abundances using
1.0 and 3.0 \kms\, showed slight trends.
Only $LS$-allowed transitions were used.
We eliminated {\bf three}
outliers after noting their transitions were not
LS-allowed, but which had not been caught by our filter.

{\bf Iron (Z = 26; $\log(Fe/N_{\rm tot}) = -4.06\pm 0.24$):}
The adopted abundance is from 98 Fe {\sc ii} lines.  Seven
Fe {\sc iii} lines give $-4.09$, while 21 Fe {\sc i} lines give
$-3.72\pm 0.37$.  In view of possible stratification
we do not consider these values.  However, with the cooler
model used in Paper II, we obtained $-3.96$, in better
agreement with Fe {\sc ii} and Fe {\sc iii},
as detailed in Paper II.  Oscillator strengths for
the first two spectra were from Fuhr and Wiese (1996).
For Fe {\sc ii}, the recent values of Mel\'{e}ndez
and Barbuy (2009)  made only a 0.01 dex
in the average abundance from Fe {\sc ii} using
transitions from Fuhr and Wiese (2006).
The iron abundance
is about a factor of three above the solar value.

{\bf Cobalt (Z = 27; $\log(Co/\Sum)\le -5.76$):} Neither
Co {\sc i} nor Co {\sc ii} is firmly identified.
An approximate upper limit is set by the absence of
the strong Co {\sc i} line $\lambda$3443.64.  If the equivalent
width were 0.5 m\AA, the abundance of Co would be $-$5.76.
The strong Co {\sc ii} line $\lambda$4062.73 gives an upper
limit $-$5.33,  assuming 0.8 m\AA.

{\bf Nickel (Z = 28; $\log(Ni/\Sum) = -6.35\pm 0.31$):}
The presence of Ni {\sc i} cannot be confirmed.
While a definite line (7 m\AA) is present within 0.01~\AA\,
of the resonance line in Multiplet 19, the {\it next}
strongest line in that multiplet is absent, as are the next
strongest {\bf four} Ni {\sc i} lines in Meggers, Corliss, and Scribner (1975).
Ni {\sc ii} is surely present, although weak.
The abundance is based on 5 lines with
measured equivalent widths from 4.3 to 10 m\AA.

{\bf Copper (Z = 29; $\log(Cu/\Sum)\le -5.81 $):}
An upper limit was set from Cu {\sc i} $\lambda$5153.24,
and equivalent width possibly $2$ m\AA.
An examination of the region of the strongest
expected lines did not confirm the identification.

{\bf Zinc (Z = 30; $\log(Zn/\Sum)\le -7.8$:} We
can only set an upper limit that is roughly solar.
An $\approx 5$ mA line measured at $\lambda^*$4911.66
near the position of Zn II $\lambda$4911.63 may be
entirely due to a Nd III line tabulated by
Ryabchikova, Ryabtsev, Kochukhov, and Bagnulo (2006).
If any Zn II feature is present, it is at the level
of the noise.

{\bf Ga through Selenium (Z = 31$-$34):}   Neither
the first nor the second spectrum of any of these elements can
be confirmed to be present.  We report an upper limit
for gallium of $-7.50$, based on an equivalent width
of $\lambda$4251.14 of 0.4 m\AA.

{\bf Bromine (Z = 35; $\log(Br/\Sum) = -6.81\pm 0.5$:}
The wavelength agreement was very good on
two HARPS spectra for the three lines with oscillator
strengths on the NIST web site.  Equivalent widths
for Br {\sc ii} $\lambda\lambda$4704.9, 4785.5, and
4816.7
of 4.1, 2.4, and 0.4 m\AA\, yield
$\log(Br/\Sum)$ of $-$6.64, $-$6.69, and $-$7.39.
We use weight 1/2 for the weakest line.
The error is
an estimate.  Br {\sc ii} is rarely
observed in late B stars, but two of these lines
were used by CH04, and all three were observed in
3 Cen A (Cowley \& Wahlgren 2006).  Br {\sc ii} is judged
weakly present in HD 65949.

{\bf Krypton (Z = 36; $\log(Kr/\Sum) = -5.85\pm 0.13$):}
Five weak Kr {\sc ii} lines on HARPS spectra have good wavelength
agreement with their laboratory values; the derived
abundances are all within a factor of two of one another.
The spectrum is weak, but present beyond
doubt.  There are no Kr {\sc ii} lines in the VALD or Kurucz
data bases.  We used transition probabilities from
the NIST site.

{\bf Rubidium (Z = 37; $\log(Ru/\Sum)\le -6.70$):}
A search
for the strongest NIST lines in the region observed did
yield some possible features for Rb {\sc ii}.  The upper limit
is based on an equivalent width of 1.2 m\AA\, for Rb {\sc ii}
$\lambda$4244.40, measured at 4244.41 on a HARPS spectrum.
The oscillator strength is from Smirnov and Shapochkin (1979).


{\bf Strontium (Z = 38; $\log(Sr/\Sum) = -6.76\pm 0.45$):}
The Sr {\sc ii} resonance lines are unmistakable, and
give abundances ranging from -6.6 to -6.2, depending on the
method of analysis (synthesis vs. equivalent width).  We
consider these lines to be affected by NLTE or stratification.
The abundance is quite uncertain.  We take it
from the subordinate lines, $\lambda\lambda$4161 and 4305.
Two very weak lines $\lambda\lambda$4312.77 and 4414.84 yield
significantly higher abundances.
We reject them as probably due to severe blending.
The oscillator strangths are from VALD, but are
not significantly different from NIST.

{\bf Yttrium (Z = 39; $\log(Y/\Sum) = -7.66\pm 0.11$):}
The
abundance is based on 14 Y {\sc ii}
lines from 3327 to 5662~\AA.
with equivalent widths from 1.3 to 18.4 m\AA.  Oscillator
strengths are from VALD.  Lines in common agree with
Fuhr and Wiese (1996).


{\bf Zirconium (Z = 40; $\log(Zr/\Sum) = -7.90\pm 0.17$):}
Our rough
estimate is based on {\bf three} Zr {\sc ii}
lines with equivalent widths
of 5.0, 9.4, and 9.6 m\AA.  Oscillator strengths from
VALD but for these lines agree sufficiently with
Ljung, et al. (2006)

{\bf Niobium (Z = 41; $\log(Nb/\Sum) = -7.26\pm 0.29$):}
The abundance
is based on 22 lines with equivalent widths from 1.6
to 20 m\AA.  The transition probabilities were
mostly taken from
Nilsson and Ivarson (2008).  Values for
$\lambda\lambda$3517 and 4119 are from VALD.
Nb {\sc ii} is not routinely identified in CP stars;
CH04 do not report an abundance for niobium
in HR 7143. Niobium
creates an odd-Z anomaly, being more abundant than
its adjacent even-Z neighbors.

{\bf Molybdenum (Z = 42; $\log(Mo/\Sum) = -7.86\pm 0.34$):}
The entry in
Table~\ref{tab:aabbs} is based on {\bf four} quite weak Mo II lines,
with one outlier weighted 1/2.  Oscillator strengths
are from Quinet (2002).

{\bf Ruthenium (Z = 44; $\log(Ru/\Sum) = 6.50\pm 0.48$):}
The abundance is based on 20 Ru {\sc ii} lines with
equivalent widths ranging from 0.3 to 22 m\AA.
We used new oscillator
strengths and partition functions recently calculated by
the Mons group  (Palmeri, et al. 2009).

{\bf Rhodium (Z = 45; $\log(Rh/\Sum) \approx -7.43$):}
The abundance is based on
only one strong Rh {\sc ii} line in Multiplet 5: $\lambda$3307.37.
The (guessed) oscillator strength, $\log(gf) = 0.00$ is from
Kurucz (1993).  CH05 attribted {\bf seven} features
to Rh {\sc ii}.  Five of these lines are below the UV cutoff of
our spectra.

{\bf Palladium (Z = 46; $\log(Pd/\Sum = -5.84\pm 0.14$):}
The result
is based on {\bf four} weak Pd {\sc i}lines, and one blended feature.
The abundances from the {\bf four} lines are within a factor
of two of one another.
Oscillator strengths are from Bi\'{e}mont, Grevesse,
and Kwiatowski (1982).
\newline\indent{\bf Silver through Tellurium: (Z = 47--52)}
\begin{itemize}
\vspace{-0.1in}
\item There is no support for Ag {\sc i} or {\sc ii}, either
from searches for the strongest lines
within our wavelength coverage.

\item We find only marginal evidence
for Cd.  The upper limit is from Cd {\sc ii}
4415.8, possibly present as as a 1 m\AA asymmetry
to the violet of a stronger, unidentified line, probably
Re {\sc ii}.

\item A search for the strongest lines yields no support
for the presence Sb {\sc ii}.

\item We derive an upper limit of
$\log(Sn/\Sum)\approx -8.42$ by assuming Sn {\sc ii} $\lambda$6453.5
has an equivalent width of 1 m\AA.  The oscillator strength
used was from NIST.

\item Tellurium: Te {\sc ii} is surely present.  Oscillator strengths
are currently being calculated, and will be reported in
due course.
\end{itemize}

{\bf Xenon (Z = 54; $\log(Xe/\Sum) = -5.42\pm 0.11$):}
The abundance is based on six Xe {\sc ii} lines with
equivalent widths from 5 to 27 m\AA.  Transition
probabilities are from
Zieli\'{n}ska, Bratasz \&
Dzier\.{z}\c{e}ga, K. (2002).
The wavelength
agreement is excellent.  The spectrum is securely
identified.

{\bf Cesium (Z = 55; $\log(Cs/\Sum)\le -7.7$):} A single, weak
feature centered at $\lambda^*$4603.78 provides an upper limit
to the Cs abundance.  The stellar feature is not broad enough
to fit the laboratory hfs (Sansonetti and Andrew 1986), though
a partial contribution from Cs II cannot be excluded.
The upper limit falls between the abundance of xenon, and
an upper limit for barium.

{\bf Barium (Z = 56; $\log(Ba/\Sum)\le -9.64$):} The
Ba {\sc ii} resonance line
$\lambda$4554 has an equivalent
width no larger than about 0.6 m\AA.  There is no indication
of the presence of the second component of the doublet,
$\lambda$4934.

{\bf Cerium (Z = 58; $\log(Ce/\Sum \le -9.79$):}
The upper limit is
based on the non appearance
of the strong Ce {\sc iii} line $\lambda$3454.39, for which we estimate
from raw UVES scans that the equivalent width cannot be
larger than about 0.1 m\AA.  Because of the wavelength placement
and intensity distribution of Ce {\sc iii}, it is more rarely
identified in CP stars.  The oscillator strength used for
$\lambda$3454 was from the DREAM site.

{\bf Praseodymium (Z = 59; $\log(Pr/\Sum) = -8.31\pm 0.21$):}
The abundance is based on 16 Pr {\sc iii} lines from
4 to 20 m\AA.  The oscillator strengths are from
the DREAM site
and line-to-line agreement is good ($\pm 0.21$ sd).
The strongest likely Pr {\sc ii} lines
are blended.  An equivalent width of 0.5 m\AA\, for $\lambda$4225
yields an upper limit of $-$7.9 for $\log(Pr/N_{\rm tot})$,
which does not seem particularly useful, since Pr {\sc iii} gives
a smaller value, and the general trend is for the third spectrum
of the lanthanides to give a {\it higher} abundance.

{\bf Neodymium (Z = 60; $\log(Nd/\Sum) = -7.03\pm 0.32$):}
The abundance
is based on 12 Nd {\sc iii} lines.
Five lines had equivalent widths over 20 m\AA,
and another three were over 10 m\AA.  Oscillator strengths
are from DREAM.

{\bf Europium (Z = 63; $\log(Eu/\Sum)\le -8.50:$):}  The
upper limit is based on Eu III
$\lambda$6666.37 (Ryabchikova, et al.
1999).  The oscillator strength is from Wyart, et al. (2008).
There is no sign of a feature on the HARPS spectrum.  A
value of 0.4 m\AA\, was used to set the upper limit.

{\bf Dysprosium (Z = 66; $\log(Dy/\Sum = -8.06\pm 0.44$):}
The abundance is based on 12 Dy {\sc iii}
lines.  All had equivalent widths $\le$ 17 m\AA.
Two lines gave abundances
about 1 dex higher than the mean of all 12 lines.  They
were averaged in, but with weight 1/2.  Since the logs were
averaged, the difference between weighting or not weighting
was only 0.1 dex.  Oscillator strengths are from DREAM.

{\bf Holmium (Z = 67; $\log(Ho/\Sum) = -8.18\pm 0.31$):}
The abundance
is based on 12 weak Ho {\sc iii} lines.
Ten of the lines used were under 10 m\AA.
Oscillator
strengths are from DREAM.

{\bf Erbium (Z = 68; $\log(Er/\Sum) = -8.80\pm 0.21$):}
The abundance is from {\bf three}
Er {\sc ii} lines, with equivalent widths between 1.4 and 3.3 m\AA.
These three are the
strongest lines by far in the Reader and Corliss (1980)
tabulation.
Oscillator strengths are from DREAM.

{\bf Ytterbium ($\log(Yb/\Sum) \approx -8.69$):} The result is
from Yb {\sc iii} $\lambda$4028.14, $W_\lambda = 0.7$~m\AA.
Yb {\sc ii}, $\lambda$4180.81 is possibly present.
$W_\lambda = 0.3$~m\AA, yields $-$9.13.  CH04 observed both
Yb {\sc ii} and Yb {\sc iii}, and obtained an
abundance from Yb {\sc iii}
0.8 dex higher than from Yb {\sc ii}.  This is qualitatively
similar to our result.

{\bf Tungsten (Z = 74; $\log(W/\Sum) \le -8.14$):}
We cannot establish the presence of W {\sc i} or W {\sc ii}.  The
W {\sc ii} line in Multiplet 1, $\lambda$3641.42 if present
is a weak feature in the wing of Ti {\sc ii} $\lambda$3641.33.
We estimate the equivalent width must be $\le$ 0.5 m\AA.
This gives an upper limit some 2.6 dex above the solar
value.  The oscillator strength used for $\lambda$3641
from VALD is 0.15 dex larger than that of Kramida and
Shirai (2006).

{\bf Rhenium (Z = 75; $\log(Re/\Sum) = -5.81\pm 0.27$):}
The Re {\sc ii}
spectrum is exceptionally well developed
in HD 65949, even though the strongest atomic lines of Re {\sc ii}
are well below our wavelength coverage.  Some 120 lines are attributed
wholly or partially to Re {\sc ii}.  New oscillator strengths
enable us to determine abundances from lines on either
side of the Balmer jump (BJ).  Using 15 lines to the violet
of the BJ, we find $-5.62\pm 0.21$; 17 lines to the red
of the BJ give $-5.97\pm 0.22$.  The sense of the difference
is that the abundance of rhenium is higher in the higher
atmosphere.  A microturbulence of 4 \kms was necessary to
remove dependence of abundance with equivalent width.  This
is reasonably attributed to hyperfine structure which is
readily visible on the HARPS spectra.  Even so, we omitted
two lines with equivalent widths of 116 and 123 m\AA.
The overall rhenium excess, is
6 dex, greater than that of any element apart
from mercury.

New oscillator strengths and
partition functions were calculated by the Mons group.
The results are presented in Appendix B1 and C1.

{\bf Osmium (Z = 76; $\log(Os/\Sum) = -5.27\pm 0.53$):}  Os {\sc ii}
is present beyond any doubt.
One can see on the high-resolution
HARPS spectra that the lines of Os {\sc ii} (and Pt {\sc ii}) are noticably
sharper than lines from lighter ions.
The abundance is based on 17 lines with equivalent
widths ranging from 5 to 38 m\AA.  The oscillator strengths
are taken from the database DESIRE, and the partition
functions are given in Appendix B1.

{\bf Platinum (Z = 78; $\log(Pt/\Sum)=-5.22\pm 0.15$):}
With Engleman's (1989) list, we identified
23 lines with Pt {\sc ii}.  There is good evidence that the
dominant isotope is $^{198}$Pt (Paper I).
We found no credible evidence for Pt I.

We adopt the absolute oscillator strength scale of
Quinet, et al. (2008, QPFB).  Only three of their lines
are available in HD 65949 (3535.89,
3551.36, and 4046.45 \AA).  Dworetsky, Story, and
Jacobs (1984, DSJ) provide oscillator strengths for
an additional {\bf six} lines, but with a different absolute
scale.  The DSJ scale was based on calculated transition
probabilities for ultraviolet lines that were used to
fix the stellar abundance of platinum in $\chi$ Lup.
Though DSJ give oscillator strengths for four lines
(see their Table II), in practice only two
($\lambda\lambda$1777.0 and 2144.0) were used for
the abundance which sets the astrophysical scale
of DSJ's Table IV.  {\bf Additionally $\lambda$4046 is
in DSJ's Table IV and QPFB.}  If we compare all four
of the common UV lines, we find the DSJ $\log(gf)'s$ are
larger by 0.30.  If we compare only the two lines
used for abundance, the corresponding figure is 0.22;
the DSJ $\log(gf)$ for $\lambda$4046 is 0.42 larger
than that of QPFB.  We have scaled down all DJS values
by 0.30 dex.  The adopted abundance is based on {\bf eight}
lines, not including $\lambda$4046, which is blended
with Hg I {and sensitive to microturbulence and isotope
shifts (Engleman 1989).  Using plausible assumptions
for the microturbulence, and isotope ratios, we can
get good agreement from $\lambda$4046 with results
from the other Pt II lines.  However, a definitive study
of isotopes is postponed to a future study.}

{\bf Gold (Z = 79; $\log(Au/\Sum)=-6.96\pm 0.52$):} The abundance
is based on Au {\sc ii}
$\lambda\lambda$4016, 4052, and 4361, with
equivalent widths of 4, 6, and 3.1 m\AA.
Respective abundances are $-$7.17, $-$7.18, and $-$6.11.
We have weighted the latter line 1/2, to form the mean
and standard deviation, assuming it is
likely a blend.
Oscillator strengths are from Rosberg
and Wyart (1997).

{\bf Mercury (Z = 80; $\log(Hg/\Sum)=-4.59\pm 0.29$):}
The strength of
Hg {\sc ii} $\lambda$3984 is
extraordinary.  The abundance of mercury used here
is an  average of four weak Hg {\sc i} and {\sc ii}
lines discussed in Papers I and II.  Oscillator strengths
from Fuhr and Wiese (1996) and Sansonetti and Reader (2001)
cause small differences from Paper II.
The uncertainty ($\pm 0.29$), is the difference,
of the averages: Hg {\sc ii} minus Hg I.

{\bf Lead (Z = 82; $\log(Pb/\Sum) \le -8.12$):}
There is no evidence for the strongest lines of
either Pb {\sc i} or {\sc ii}.  The upper limit used here assumes
an equivalent width of 0.2 m\AA\, for Pb {\sc ii} $\lambda$5042.
If $\log(Pb/N_{\rm tot})$ were as large as $-$6.0,
$\lambda$5042 would have an equivalent width of some
16 m\AA, and be easily detected.  It is clear that
lead is significantly lower
in abundance than osmium, platinum, or mercury.

{\bf Bismuth (Z = 83; $\log(Bi/\Sum)=-8.0\pm 0.5$):}  An
upper limit is from the strongest lines,
$\lambda\lambda$4079 and 5209,
discussed by
Dolk, Litz\'{e}n, \& Wahlgren (2002, DLW) for HR 7775.
There is broad, weak absorption near the position of the
$\lambda$5209 components, but the Bi II hfs components
do not fit it well.  A measured feature, at $\lambda^*$4259.46
is too far from the laboratory position.  The upper limit is
based on a synthesis that assumes a contribution from Bi II
at the level of the noise.

{\bf Thorium (Z = 90; $\log(Th/\Sum)=-9.14\pm 0.17$)}
The abundance is based on {\bf eight} lines with measured equivalent
widths from 1.0 to 4.6 m\AA.  Oscillator strengths are
from DREAM.  {\bf Partition functions for Th {\sc ii}
and {\sc iii} were calculated from energy levels of
Blaise and Wyart (2009).  Results differ only slightly
from values used at Michigan for several decades.}


\section{New Partition Functions}
\label{app:pfn}
Partition functions can be a significant source of
error for stellar abundances if they are inaccurate.
In most of the present work,
partition functions were calculated from published atomic energy
levels (e.g. Moore 1949-1958), or levels produced by
the Cowan (1981) atomic structure code
(cf. Cowley and Barisciano 1994).  The present work uses
new partition functions for the
first through third spectra of ruthenium, rhenium,
and osmium (see Table \ref{tab:pfn}).  These were
calculated on the basis of the experimental energy levels
available in the literature adding, in each case, additional
theoretical values deduced from HFR calculations.  Relevant
references are indicated by footnotes to the table.  Full
citations appear among the main references.

\begin{table*}
\begin{minipage}{126mm}

\caption{New partition functions for Ru,
 Re and Os atoms and ions.
 \label{tab:pfn}}

\begin{tabular}{@{}ccccccccccccccccccc}\hline

T (K){\Rv}  & &Ruthenium&& & \multicolumn{2}{l}{Rhenium}&&
\multicolumn{2}{l}{Osmium}   \\
&Ru I$^a$ &  Ru II$^b$&   Ru III$^b$ &Re I$^c$&  Re II$^d$& Re III$^e$
&Os I$^f$&Os II$^g$&Os III$^h$ \\  \hline

3000 &22.33 &17.13& 16.47 &6.10  &7.03 &6.02  &12.80& 13.40& 11.85\\
3500 &24.92 &18.65 &17.40 &6.27 &7.08 &6.05 & 14.28& 14.79& 12.87\\
4000 &27.72& 20.25& 18.27 &6.55 &7.18 &6.12&  15.93& 16.31& 13.93\\
4500 &30.75 &21.94 &19.13  &6.98 &7.36 &6.24& 17.75& 17.97& 15.02\\
5000 &34.01 &23.73& 20.03  &7.58 &7.63 &6.43&  19.74& 19.77& 16.15\\
5500 &37.49 &25.62 &20.98  &8.38 &8.01 &6.69&  21.92& 21.70& 17.33\\
6000 &41.20 &27.60 &21.99  &9.39 &8.53 &7.04&24.29& 23.77& 18.56\\
6500 &45.15 &29.68 &23.09  &10.62 &9.18 &7.48& 26.87& 25.98& 19.85\\
7000 &49.37 &31.85 &24.25  &12.10 &9.98 &8.02&  29.65& 28.33& 21.22\\
7500 &53.87 &34.10 &25.49  &13.83  &10.95 &8.67&  32.67& 30.83& 22.66\\
8000 &58.70 &36.45 &26.80  &15.84  &12.20  &9.41 &  35.93& 33.46& 24.18\\
8500 &63.90 &38.87 &28.17  &18.15 &13.41 &10.26&  39.45& 36.24& 25.78\\
9000 &69.53 &41.39 &29.60  &20.78 &14.91 &11.22&  43.27& 39.16& 27.47\\
9500 &75.64 &43.98 &31.08  &23.76 &16.58 &12.27&  47.41& 42.21& 29.23\\
10000& 82.30 &46.66 &32.61  &27.11 &18.45 &13.43&  51.90& 45.41& 31.08\\
10500& 89.58 &49.42 &34.18  &30.88 &20.50 &14.68&  56.79& 48.73& 33.01\\
11000& 97.55 &52.27 &35.79  &35.20 &22.74 &16.03&  62.10& 52.19& 35.03\\
11500& 106.29 &55.21 &37.43  &39.79 &25.16 &17.48&  67.89& 55.79& 37.12\\
12000& 115.88 &58.23 &39.10  &45.01 &27.78 &19.01&  74.19& 59.53& 39.28\\
12500& 126.39 &61.35 &40.81  &50.80 &30.58 &20.64&  81.07& 63.40& 41.53\\
13000& 137.89 &64.56 &42.54  &57.20 &33.58 &22.34&  88.55& 67.41& 43.84\\
13500& 150.45 &67.87 &44.29  &64.25 &36.77 &24.14&  96.69& 71.56& 46.23\\
14000& 164.15 &71.28 &46.08  &71.99 &40.16 &26.01&  105.55& 75.86& 48.69\\
14500& 179.06 &74.81 &47.88  &80.48 &43.75 &27.96&  115.16& 80.31& 51.21\\
15000& 195.22 &78.45 &49.71  &89.75 &47.53  &29.98&  125.58& 84.91& 53.80\\
15500& 212.70 &82.22 &51.56  &99.85 &51.52  &32.08&  136.84& 89.67& 56.46\\
16000& 231.54 &86.12 &53.44  &110.82 &55.73 &34.25&  148.99& 94.60& 59.18\\
16500& 251.81 &90.16 &55.34  &122.69 &60.14 &36.49&  162.08& 99.71& 61.96\\
17000& 273.53 &94.34 &57.27  &135.50 &64.78 &38.79& 176.13& 104.99& 64.80\\
17500& 296.75 &98.69 &59.22  &149.30 &69.63 &41.16&  191.18& 110.46& 67.71\\
18000& 321.50 &103.22 &61.20  &164.10 &74.73 &43.59&  207.27& 116.12& 70.67\\
18500& 347.80 &107.92 &63.20  &179.95 &80.05  &46.20&  224.41& 121.98& 73.70\\
19000& 375.69 &112.82 &65.24  &196.86 &85.63 &48.64&  242.65& 128.06& 76.78\\
19500& 405.17 &117.94 &67.30  &214.87 &91.46 &51.25&  261.99& 134.35& 79.92\\
20000& 436.27 &123.27 &69.39  &233.99 &97.54  &53.92&  282.46& 140.88& 83.13\\
20500& 469.00 &128.85 &71.52  &254.24 &103.90& 56.64&  304.07& 147.64& 86.39\\
21000& 503.35 &134.68 &73.68  &275.64 &110.53 &59.42&  326.84& 154.64& 89.71\\
21500& 539.34 &140.78 &75.87  &298.20 &117.44  &62.25&  350.76& 161.91& 93.20\\
22000& 576.97 &147.16 &78.10  &321.93 &124.65  &65.13&  375.86& 169.43& 96.53\\
22500& 616.23 &153.85 &80.37  &346.84 &132.16 &68.07&  402.12& 177.23& 100.03\\
23000& 657.11  &160.86 &82.68  &372.93 &139.98 &71.05&  429.56& 185.31& 103.60\\
23500& 699.62  &168.20  &85.02  &400.21 &148.12 &74.08&  458.17& 193.69& 107.22\\
24000& 743.73 &175.90  &87.42   &428.68 &156.59 &77.16&  487.95& 202.36& 110.90\\
24500& 789.44 &183.97  &89.86  &458.32 &165.39 &80.28&  518.88& 211.34& 114.65\\
25000& 836.73 &192.43  &92.35  &489.16 &174.53& 83.45&  550.96& 220.64& 118.46\\
25500& 885.58 &201.30  &94.89  &521.16 &184.02 &86.67&  584.18& 230.26& 122.34\\
26000& 935.97 &210.60  &97.48  &554.33 &193.87 &89.93&  618.53& 240.21& 126.27\\

\hline
\end{tabular}
\end{minipage}
\end{table*}

\begin{table*}
\begin{minipage}{126mm}
\contcaption{}
\begin{tabular}{@{}ccccccccccccccc}\hline

T (K){\Rv}& &Ruthenium&& & \multicolumn{2}{l}{Rhenium}&
& \multicolumn{2}{l}{Osmium}   \\
&Ru I$^a$ &  Ru II$^b$&   Ru III$^b$ &Re I$^c$&  Re II$^d$& Re III$^e$&Os I$^f$&Os II$^g$
&Os III$^h$ \\ \hline
26500& 987.89 &220.34 &100.13&  588.66& 204.08& 93.23&  653.99& 250.51& 130.27\\
27000& 1041.31 &230.55 &102.83&  624.13& 214.67& 96.57&  690.55& 261.15& 134.34\\
27500& 1206.20 &241.23 &105.60&  660.74& 225.63& 99.95 & 728.18& 272.15& 138.47\\
28000& 1152.56 &252.42 &108.43&  698.46& 236.98& 103.37&  766.88& 283.51& 142.66\\
28500& 1210.33 &264.12 &111.32&  737.29& 248.71& 106.83&  806.62& 295.23& 146.93\\
29000& 1269.52 &276.36 &114.29&  777.21& 260.85& 110.33&  847.38& 307.33& 151.25\\
29500& 1330.08 &289.15 &117.32&  818.20& 273.38& 113.87&  889.15& 319.80& 155.65\\
30000& 1391.98 &302.51 &120.43&  860.24& 286.32& 117.44&  931.89& 332.66& 160.11\\
30500& 1455.21 &316.46 &123.62&  903.32& 299.67& 121.04&  975.59& 345.91& 164.63\\
31000& 1519.73 &331.00 &126.89&  947.41& 313.43& 124.68&  1020.22& 359.55& 169.23\\
31500& 1585.52 &346.17 &130.25&  992.49& 327.61& 128.35&  1065.77& 373.58& 173.89\\
32000& 1652.54 &361.98 &133.69&  1038.55& 342.21& 132.06&  1112.20& 388.01& 178.61\\
32500& 1720.76 &378.43 &137.23&  1085.56& 357.23& 135.79&  1159.49& 402.85& 183.41\\
33000& 1790.16 &395.55 &140.86&  1133.50& 372.68 & 139.56&  1207.62& 418.20& 188.27\\
33500& 1860.71 &413.35 &144.59&  1182.34& 388.55& 143.35&  1256.57& 433.73& 193.20\\
34000& 1932.37 &431.85 &148.42&  1232.08& 404.85& 147.17&  1306.30& 449.79& 198.19\\
34500& 2005.12 &451.05 &152.36&  1282.68& 421.58& 151.02&  1356.81& 466.25& 203.25\\
35000& 2078.93 &470.98 &156.41&  1334.12& 438.74&  154.89&  1408.05& 483.13& 208.38\\

\hline
\end{tabular}
\\
$^a$	Ru I : experimental levels completed with HFR values
as described in Fivet, et al. (2009). \\
$^b$	Ru II-III : experimental levels completed with HFR values
as described in Palmeri, et al. (2009). \\
$^c$	Calculated using the experimental energy levels
of Klinkenberg, et al. (1957)
with \\additional semi-empirical HFR values from Palmeri,
et al. (2006).\\
$^d$	Calculated using the experimental energy levels of
Meggers, et al. (1958),
Wyart (1977)\\ and Wahlgren, et al. (1997) with additional
semi-empirical HFR values from \\Palmeri, et al. (2005).\\
$^e$	Calculated using the HFR energy levels predicted
in the present work.\\
$^f$	Experimental levels with additional HFR values
taken from Quinet, et al. (2006).\\
$^g$	Experimental levels completed with HFR values
as described in Quinet, et al. (2006).\\
$^h$	Experimental levels (unpublished) kindly communicated
by A.N. Ryabtsev (2009) with \\additional HFR values.\\

\end{minipage}
\end{table*}

\section{Transition probabilities in Re II}
\label{app:loggf}

Transition probabilities had been obtained by Palmeri
et al. (2005) for 45 lines of Re II as a part of
the general project to build the
{\bf D}atabasE for th{\bf e SI}xth {\bf R}ow
{\bf E}lements (DESIRE, Fivet,
et al. 2007).  They had used a
combination of theoretical branching fractions with 
radiative lifetimes measured by time-resolved 
laser-induced fluorescence spectroscopy. The results 
reported were for transitions depopulating the levels 
with measured lifetimes. Using the same relativistic 
Hartree-Fock method, including core-polarization 
effects, the sample of results obtained by these 
authors has been considerably extended in the present 
study. More precisely, in the physical model used, the 
interactions between the $5d^5ns$ ($n=6-8$), $5d^46sns$
($n=6-8$), $5d^6$, $5d^56d$, $5d^46s6d$, $5d^36s^26d$,
$5d^46p^2$ and
$5d^36s6p^2$ (even parity) and $5d^5np$ ($n=6-8$), $5d^46snp$
($n=6-8$), $5d^36s^26p$ (odd parity) configurations were
retained. A least-squares fitting of the calculated 
eigenvalues of the hamiltonian to the observed energy 
levels was applied, using experimental levels from 
Meggers et al. (1958), Wyart (1977), and Wahlgren et 
al. (1997). We retained 44 even-parity and 55 
odd-parity {\bf levels in} the fit leading to standard
deviations of 135 (even) and 192 cm$^{-1}$ (odd levels),
respectively. The transition probabilities and 
oscillator strengths of the strongest ($\log{gf} >
-1.0$) transitions of Re II with $\lambda > 2000$~\AA\, are
reported.  Additionally, lines identified wholly or
partially as Re II in HD 65949 are included.

\begin{table*}
\caption{\label{ReIIgf} New oscillator strengths for Re II
transitions.  Only transitions with
$\lambda > 2000$~\AA\, are included in the table.}
\begin{tabular}{cccc||cccc||cccc}      \hline
$\lambda$(\AA){\Rv}& Lower$^a$ &$\log{gf}^b$&CF$^c$&
$\lambda$(\AA){\Rv}& Lower$^a$ &$\log{gf}^b$&CF$^c$&
$\lambda$(\AA){\Rv}& Lower$^a$ &$\log{gf}^b$&CF$^c$\\
  \hline
2009.926& 19140 &-0.63&  0.093&2449.038& 18846 &-0.44& 0.198&3523.160&31013 & -2.01& 0.007  \\
2018.547& 14883 &-0.87&  0.139&2455.836& 14352 &-0.82& 0.066&3527.112&36064 & -1.76& 0.014  \\
2023.652& 14883 & 0.00&  0.568&2467.567& 14931 &-0.68&-0.140&3542.724&36064 & -1.51&-0.026  \\
2027.206& 20463 &-0.89& -0.063&2468.475& 22545 &-0.95& 0.041&3580.134&17224 & -0.64& 0.268  \\
2042.642& 13777 &-0.83&  0.099&2469.389& 18846 &-0.93& 0.072&3581.422&24763 & -2.17& 0.005  \\
2053.603& 20463 &-0.49& -0.191&2470.610& 20463 &-0.66&-0.068&3601.602&33169 & -1.95&-0.022  \\
2055.255& 14824 &-0.75& -0.113&2471.050& 19140 &-0.86&-0.044&3609.339&27746 & -2.51& 0.002  \\
2059.765& 14931 &-0.92& -0.044&2475.186& 23894 &-0.66&-0.102&3626.781&26237 & -2.24&-0.009  \\
2064.163& 18846 &-0.81&  0.102&2478.992& 14824 &-0.89& 0.137&3647.530&32258 & -1.55&-0.014  \\
2075.134& 19140 &-0.88& -0.040&2487.449& 19140 &-0.88& 0.058&3656.830&32258 & -2.07&-0.007  \\
2075.720& 14883 &-0.73&  0.073&2489.028& 28095 &-0.50&-0.155&3697.925&26768 & -1.79& 0.031  \\
2083.695& 14883 &-0.26&  0.300&2490.200& 20782 &-0.42&-0.233&3714.509&30225 & -2.27&-0.005  \\
2085.777& 14931 &-0.28& -0.178&2502.350& 20976 &-0.06&-0.205&3731.675&32876 & -1.61&-0.020  \\
2091.547& 20463 &-0.45&  0.171&2550.086& 20463 &-0.65&-0.126&3773.011&33169 & -3.51& 0.000  \\
2091.932& 17224 &-0.94& -0.067&2553.525& 23894 &-0.81&-0.046&3782.963&33169 & -2.31& 0.005  \\
2108.934& 18846 &-0.91& -0.121&2553.604& 17224 &-0.83&-0.116&3783.779&30718 & -2.43& 0.003  \\
2111.866& 19140 &-0.41&  0.117&2554.628& 20463 &-0.18&-0.200&3791.586&29077 & -2.38& 0.003  \\
2114.251& 20976 &-0.60& -0.067&2557.419& 25321 &-0.91&-0.076&3800.964&18846 & -1.71&-0.128  \\
2133.120& 20463 &-0.89&  0.052&2566.374& 28095 &-0.64&-0.110&3823.636&37319 & -2.03&-0.014  \\
2134.792& 18846 &-0.86& -0.090&2568.636& 14883 &-0.45& 0.218&3826.548&31013 & -2.49& 0.002  \\
2144.083& 22545 &-0.74& -0.088&2571.802& 14931 &-0.64& 0.221&3830.551&23341 & -1.90&-0.020  \\
2145.896& 20463 &-0.73& -0.092&2576.236& 26768 &-0.80& 0.059&3839.540&31013 & -1.85&-0.012  \\
2170.806& 20463 &-0.73&  0.073&2588.517& 29639 &-0.98&-0.046&3847.742&29077 & -2.00&-0.012  \\
2172.108& 23146 &-0.63&  0.102&2588.578& 20976 &-0.90&-0.106&3858.509&26768 & -2.29& 0.006  \\
2177.587& 20463 &-0.85&  0.063&2608.497& 14352 &-0.40& 0.214&3873.489&37319 & -1.89& 0.026  \\
2181.779& 17224 &-0.72&  0.074&2610.112& 30983 &-0.89&-0.036&3915.407&37382 & -2.34& 0.005  \\
2187.911& 17224 &-0.84&  0.070&2610.541& 25988 &-0.96& 0.060&3939.368&29773 & -2.14& 0.019  \\
2190.260& 14883 &-0.60&  0.222&2611.537& 24763 &-0.95&-0.038&3964.111&30225 & -2.06&-0.008  \\
2195.273& 20976 &-0.70& -0.081&2616.718& 18846 &-0.84&-0.087&3984.242&18846 & -2.01&-0.137  \\
2197.126& 20976 &-0.91&  0.050&2635.838& 17224 &-0.67& 0.128&4020.856&36064 & -1.65& 0.028  \\
2214.275&     0 & 0.04&  0.406&2637.006& 19140 &-0.78&-0.068&4031.464&19140 & -2.11& 0.041  \\
2216.157& 20463 &-0.72&  0.094&2731.566& 18846 &-0.68&-0.128&4032.355&32258 & -2.02& 0.008  \\
2229.106& 21629 &-0.75&  0.084&2733.030& 17224 &-0.35& 0.219&4042.758&34937 & -2.08& 0.007  \\
2247.555& 20463 &-0.92&  0.046&2750.551& 30983 &-0.71& 0.068&4089.913&26237 & -2.66& 0.073  \\
2248.627& 14931 &-0.86&  0.070&2813.528& 30983 &-0.95&-0.078&4091.972&31013 & -1.79&-0.012  \\
2248.763& 25321 &-0.92&  0.054&2875.720& 28095 &-0.94&-0.060&4120.373&32876 & -2.39& 0.004  \\
2261.871& 18846 &-0.83&  0.061&3103.166& 17224 &-0.96&-0.163&4135.441&32876 & -2.06& 0.009  \\
2272.645& 19140 &-0.98&  0.069&3105.075& 36064 &-0.93&-0.058&4152.688&29728 & -1.91&-0.024  \\
2275.253&     0 &-0.04&  0.402&3298.542& 36064 &-1.84& 0.009&4236.149&29077 & -2.54& 0.004  \\
2286.614& 19140 &-0.74& -0.058&3299.805& 24763 &-1.77& 0.016&4240.174&30225 & -3.09& 0.002  \\
2295.214& 25321 &-0.18&  0.251&3303.213& 14883 &-0.95& 0.123&4299.903&29427 & -2.09& 0.023  \\
2298.100& 20782 &-0.12&  0.278&3317.743& 22545 &-1.64& 0.021&4311.697&32258 & -1.97& 0.009  \\
2301.603& 23894 &-0.97&  0.031&3318.789& 25321 &-1.07& 0.072&4330.674&30718 & -1.82&-0.024  \\
2301.805& 20976 &-0.78&  0.094&3331.309& 37319 &-1.12& 0.054&4356.283&29728 & -2.41& 0.006  \\
2303.985& 23341 &-0.64& -0.116&3338.574& 30983 &-1.07&-0.080&4380.967&30983 & -2.00& 0.027  \\
2308.444& 20976 &-0.78& -0.125&3360.876& 33169 &-2.67&-0.002&4409.547&26768 & -2.43& 0.010  \\
2324.430& 23722 &-0.87&  0.105&3365.390& 26666 &-2.23& 0.007&4422.999&22545 & -2.16& 0.025  \\
2336.928& 21629 &-0.46& -0.128&3379.078& 14352 &-1.09& 0.170&4452.657&30225 & -2.06&-0.012  \\
2360.297& 23894 &-0.74& -0.070&3395.649& 30225 &-2.01& 0.006&4481.343&21629 & -2.07&-0.024  \\
2368.563& 23894 &-0.56&  0.098&3403.707& 30225 &-1.33& 0.031&4520.959&34937 & -1.98&-0.011  \\
2370.765& 14883 &-0.84&  0.098&3407.780& 23341 &-1.77&-0.019&4584.473&23341 & -2.50&-0.006  \\
2373.461& 14931 &-0.95& -0.042&3411.501& 27746 &-2.29&-0.003&4904.356&24763 & -2.49& 0.006  \\
2378.510& 27746 &-0.88&  0.042&3427.961& 30225 &-2.72&-0.002&4909.738&29077 & -2.48& 0.012  \\
2382.075& 24763 &-0.73&  0.092&3433.839& 14824 &-1.96&-0.081&5286.696&26237 & -2.83&-0.009  \\
2386.899& 20976 &-0.33&  0.201&3434.898& 30225 &-2.63& 0.002&  \\
2389.609& 21629 &-0.89&  0.058&3446.447& 14931 &-1.86&-0.024&  \\
2403.034& 23341 &-0.80&  0.067&3452.658& 28095 &-2.22&-0.005&  \\
2418.195& 25988 &-0.33&  0.127&3473.423& 37319 &-1.90& 0.016&  \\
2418.394& 24763 &-0.64&  0.084&3485.352& 30983 &-2.14& 0.005&  \\
2421.405& 21629 &-0.96& -0.041&3486.190& 26768 &-1.44&-0.020&  \\
2433.732& 26237 &-0.54& -0.100&3497.692& 32345 &-1.99& 0.010&  \\
\hline
   \end{tabular}
$^a$Experimental levels taken from Meggers et al. 1958;
Wyart 1977; Wahlgren et al. 1997\\

$^b$HFR+CPOL calculations. Fore more details, see
Palmeri et al. (2005).\\

$^c$Cancellation factor as defined by Cowan (1981).\\
   \end{table*}

\label{lastpage}
\end{document}